# Modeling the residual queue and queue-dependent capacity in a static traffic assignment problem

Hao Fu[a], William H.K. Lam[b,*], Wei Ma[b,*], Yuxin Shi[b],
Rui Jiang[a], Huijun Sun[a], Ziyou Gao[a]

[a] *School of Systems Science, Beijing Jiaotong University, Beijing, China*
[b] *Department of Civil and Environmental Engineering, The Hong Kong Polytechnic University, Hong Kong, China*

* Corresponding author, E-mail address: wei.w.ma@polyu.edu.hk (W. Ma); william.lam@polyu.edu.hk (WHK. Lam)

**Abstract**

The residual queue during a given study period (e.g., peak hour) is an important feature that should be considered when solving a traffic assignment problem under equilibrium for strategic traffic planning. Although studies have focused extensively on static or quasi-dynamic traffic assignment models considering the residual queue, they have failed to capture the situation wherein the equilibrium link flow passing through the link is less than the link physical capacity under congested conditions. To address this critical issue, we introduce a novel static traffic assignment model that explicitly incorporates the residual queue and queue-dependent link capacity. The proposed model ensures that equilibrium link flows remain within the physical capacity bounds, yielding estimations more aligned with data observed by traffic detectors, especially in oversaturated scenarios. A generalized link cost function considering queue-dependent capacity, with an additional queuing delay term is proposed. The queuing delay term represents the added travel cost under congestion, offering a framework wherein conventional static models, both with and without physical capacity constraints, become special cases of our model. Our study rigorously analyzes the mathematical properties of the new model, establishing the theoretical uniqueness of solutions for link flow and residual queue under certain conditions. We also introduce a gradient projection-based alternating minimization algorithm tailored for the proposed model. Numerical examples are conducted to demonstrate the superiority and merit of the proposed model and solution algorithm.

## 1 Introduction

### 1.1 Background

The residual queue during a given study period (e.g., peak hour) is an important feature that should be considered when solving a traffic assignment problem under equilibrium for strategic policy planning. Because a residual queue occupies space in the road segment, the link capacity (i.e., link exit capacity), defined as the number of vehicles that can pass a link during a predefined time period, is not always equal to the link physical capacity (i.e., the maximum flow). In other words, the link capacity is influenced by the residual queue under congested (i.e., oversaturated or hypercongested) conditions. Under such oversaturation, this queue-dependent link capacity results in an equilibrium link flow less than or equal to the link physical capacity. As shown in Figure 1, the majority of traffic flows observed by traffic detectors are less than the link physical capacity, regardless of the degree of congestion.

Although studies have focused extensively on static traffic assignment models that consider the residual queue (Lam and Zhang, 2000; Bliemer et al., 2014; Bliemer and Raadsen, 2020), they have failed to capture the situation wherein the equilibrium link flow is less than the link physical capacity under congested conditions. Traditional static models make several assumptions that result in overestimation of equilibrium link flows, especially under congested conditions (e.g., estimated traffic flows equaling or exceeding the link physical capacity, as shown by the black dot in Figure 1) (Bliemer et al., 2014). As a result, the link flows estimated using a static traffic assignment model might not match the link flows observed by traffic detectors in the presence of traffic queues on the roads (i.e., the blue circles in Figure 1). The queue-dependent capacity proposed in this paper is approximated by the lower-half green dashed line in Figure 1, while the vertical black dashed line represents the fixed physical capacity.

Therefore, a new static traffic assignment model that explicitly considers the residual queue and queue-dependent link capacity is proposed in this paper. The estimated link flow obtained from the proposed model even under congested conditions can be less than or equal to the link physical capacity. Accordingly, the estimated link flow would be consistent with the flow observed by a traffic detector. The residual queue and corresponding queuing delay can also be obtained from the proposed model.

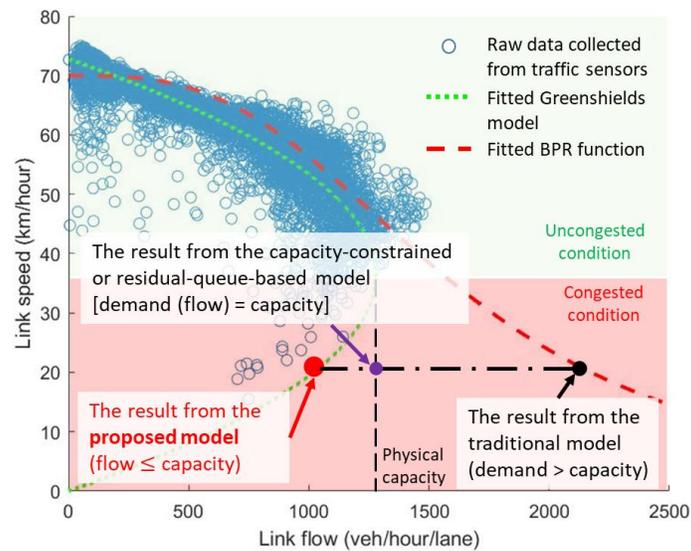

Figure 1. Difference between the results obtained with the traditional and proposed models

## 1.2 Motivations and contributions

Referring to Figure 1, the uncongested/congested condition in this paper represents the situations when travel speed is larger/smaller than the speed at physical capacity. Figure 1 uses observed data from traffic detectors to show that the link flows under congested conditions are less than the link physical capacity. However, the link flows estimated using traditional static traffic assignment models may be larger than the link physical capacity and thus cannot completely reflect the actual traffic situation under congested conditions. The model proposed in this paper aims to address this inconsistency so that the link flows estimated by the proposed model under congested conditions are consistent with the observed data.

The motivation behind this paper can be clearly demonstrated through an example and comparison of the results of the proposed model with those of some conventional models. The toy network shown in Figure 2, which includes four nodes and four links, has been used in multiple papers to demonstrate motivation (Bliemer et al., 2014; Lam and Zhang, 2000). In this network, one OD pair ($r \rightarrow s$) and two different paths are connected by links {1,2,4} and {1,3,4}. A fixed travel demand of 1200 vehicles/hour (veh/hr) and a period of 1 hour are used in this example. The free flow travel time and link physical capacity are shown in Figure 2.

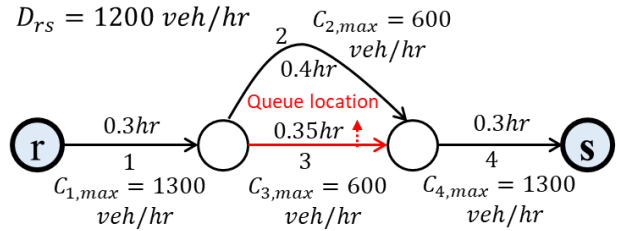

Figure 2. Demonstration of motivation using a toy network

Table 1. Comparison of the results pertaining to link 3 from different traffic assignment models

| Models | Link flow (veh/hr) | Link capacity (veh/hr) | Residual queue (veh) | Queuing delay (hr) | Total link travel time (hr) |
|---|---|---|---|---|---|
| Traditional unconstrained model | 676 | 600 | \ | \ | 0.435 |
| Fixed capacity constraint (Nie et al., 2004) | 600 | 600 | \ | 0.058 | 0.460 |
| Fixed capacity constraint with residual queue (Lam and Zhang, 2000) | 600 | 600 | 60 | 0.013 | 0.440 |
| Flow-dependent capacity constraint (Bliemer et al., 2014; Bliemer and Raadsen, 2020) | 600 | 600 | 0 | \ | \ |
| Queue storage capacity constraint (Smith et al., 2019) | 600 | 600 | 34.5 | 0.058 | 0.460 |
| Queue-dependent capacity constraint (this paper) | 573 | 573 | 54 | 0.047 | 0.450 |

As Link 3 is the bottleneck link in this toy network, the results pertaining to Link 3 from different traffic assignment models are presented in Table 1. For the conventional static traffic assignment model without a capacity constraint (Traditional unconstrained model in Table 1), the link flow under equilibrium on Link 3 is 676 veh/hr, which exceeds the link physical capacity. For the model with a fixed capacity constraint (Nie et al., 2004), the link flow on Link 3 (600 veh/hr) is capped at its physical capacity. To reach equilibrium in this model, an extra delay is considered by using the Lagrange multiplier of the optimization problem and the delay is added to the Bureau of Public Roads (BPR) function to evaluate the path travel time.

In the model with a fixed capacity constraint and a residual queue (Lam and Zhang, 2000), a flow of 600 veh/hr can pass Link 3 during the considered period, and a residual queue of 60 vehicles is retained on this link (Lam and Zhang, 2000). Interestingly, the flow-dependent capacity-constrained model described in Bliemer et al. (2014) and Bliemer and Raadsen (2020) assumes that a queue forms on Link 1, before the bottleneck link (Link 3). This assumption leads to the estimation that the queue on Link 1 contains 600 vehicles after 1 hour (Bliemer et al., 2014; Bliemer and Raadsen, 2020), which in turn is due to the assumption that the link capacity is dependent on the downstream links according to the first-order node model used in their study. Under static user equilibrium, all travelers choose the bottom path connected by links {1,3,4}, such that Link 3 is the bottleneck and only 600 vehicles can pass Link 1. In the spatial queuing model with a queue storage capacity constraint proposed by Smith et al. (2019), a residual queue of 34.5 veh/hr is retained on Link 3, which can spill back to upstream link 1 if demand increases significantly.

However, none of these models have been able to replicate the observation that the flow on the bottleneck link under congested conditions is less than the link physical capacity (see the blue circles in Figure 1). The proposed model can take account of this below-capacity flow in congested conditions, whereby in the presence of a traffic queue (i.e., congestion), the link capacity is reduced from 600 veh/hr (i.e., the link physical capacity) to 573 veh/hr. In other words, only 573 vehicles can pass through the bottleneck link per hour (Link 3), and 54 vehicles are retained in the queue on Link 3 after 1 hour. It is observed that the residual queue locates on bottleneck (link 3) in the proposed model and Lam and Zhang (2000), while on link 1 in Bliemer et al. (2014). This difference comes from that in this paper the link (exit) capacity is implicitly considered much smaller than the link entrance capacity due to the drivers' gaze behavior at intersection or traffic control strategies (Wong et al., 2007; Ringhand et al., 2022). We also find that the total travel time over the whole network in the proposed model and Bliemer et al. (2014) is 1.56 hr and 1.85 hr, respectively. This finding could stimulate a new traffic management strategy for future research, based on the idea of Braess paradox, that a bottleneck can be manually installed at the exit of link 3 to ensure the queue location so that the total travel time over the network can be reduced.

Using the above example as motivation, conventional static traffic assignment models have been shown to be incapable of generating practical traffic flows on links, regardless of the capacity constraint condition. Thus, a new static traffic assignment model that explicitly considers the queue-dependent link capacity and residual queue is proposed herein with the aim of producing more realistic link flows that can match the data collected by traffic detectors. The contributions of this paper can be summarized as below:

- A new static traffic assignment model that explicitly considers the **residual queue** is proposed to ensure that the assigned link flows are less than or equal to the link physical capacity and coincide with the flows observed by traffic detectors, especially under congested conditions.

- A generalized link cost function with a **queue-dependent capacity** is proposed, and a queuing delay term is introduced to represent the additional travel cost under congested conditions. Using this generalized link cost function, conventional static models with or without a physical capacity constraint can be considered special cases of the proposed model.

- The mathematical properties of the proposed model are analyzed to better characterize the newly formulated model. In particular, the **uniqueness** of the solutions with respect to link flow and residual queue can be proven theoretically under certain conditions.

- An efficient solution algorithm, gradient projection – alternating minimization (GP-AM), is

developed to solve the proposed static traffic assignment model and guarantee the convergence of the algorithm.

## 1.3 Literature review

The concept of traffic assignment, which is a crucial process in the planning and operation of transportation systems, has been extensively studied for decades (Wardrop, 1952; Shao et al., 2006; Ordónez and Stier-Moses, 2010; Ma and Qian, 2017; Xu et al., 2023; Xu et al., 2024). At its core, the traffic assignment problem involves allocating a given set of trip interchanges to a defined network in a way that is consistent with the travel behavior of the users. Over the years, many models have been proposed to understand and manage the complexities of the problem (Sheffi, 1985; Bliemer et al., 2017; Xie and Nie, 2019; Xu et al., 2021; Xu et al., 2022).

The static traffic assignment model, a traditional approach, assumes that the travel demand and network conditions are homogeneous over a defined period (Beckmann et al., 1956). As summarized in Table 2, although this model has been used extensively, it has been criticized for its inability to accurately depict traffic flows under congested conditions by hydrodynamic traffic flow theory (Bell and Iida, 1997; Bliemer et al., 2017; Raadsen and Bliemer, 2019b). Specifically, the observed traffic flow under congested conditions is less than the link physical capacity.

Table 2. Categories of static (quasi-dynamic) traffic assignment models

| Literature | Queuing delay | Queue | Spillback | Below-capacity flow under congested conditions | Solution properties | |
|---|---|---|---|---|---|---|
| | | | | | Unique link queue | Unique link flow |
| Nie et al. (2004) | ✓ | | | | | ✓ |
| Lam et al. (2000) | ✓ | ✓ | | | - | - |
| Bliemer et al. (2014) | ✓ | ✓ | | | | |
| Smith et al. (2019) | ✓ | ✓ | ✓ | | | |
| Bliemer and Raadsen (2020) | ✓ | ✓ | ✓ | | | |
| This paper | ✓ | ✓ | | ✓ | ✓ | ✓ |

To tackle the discrepancies in traffic flow under congested conditions, researchers ventured into incorporating physical capacity constraints into their models. This idea was conceived as a countermeasure to the persistent issue of the assigned traffic flow surpassing the physical capacity of the link (e.g., Yang and Yagar, 1995; Nie et al., 2004). Yang and Yagar (1995) adopted the capacitated traffic assignment into a bilevel model to optimize signal timing in a saturated road network. Nie et al. (2004) investigated the static traffic assignment with physical capacity constraints and developed an augmented Lagrangian multiplier approach and an inner penalty function approach to solve this problem. However, the residual queue, which occur frequently in reality, has not been explicitly considered in these static traffic assignment models with a physical capacity constraint.

To describe the relationship between link travel time and link flow in traffic assignment problems, various link cost functions (also referred to as volume delay functions, link performance functions, etc.) have been developed in previous studies. These functions can generally be classified into three categories (Wu et al., 2021; Pan et al., 2024): flow-based, density-based, and queue-based methods as shown in Table A.1 in Appendix A. Flow-based methods are the most commonly used in static traffic assignment models. For example, based on the classical BPR function, Davidson (1966) developed a flow – travel time relationship for use in traffic planning. Akçelik (1991) further extended this relationship by proposing a time dependent form using coordinate transformation technique. Spiess (1990) introduced a conical link cost function to characterize the specific congestion behavior of a road link, while Small (1983) proposed a piecewise link cost function that distinguishes between flow conditions less than (or equal to) and greater than physical capacity. To more realistically reproduce actual speed under hypercritical conditions, Kucharski and Drabicki (2017) proposed a density-based link cost function which is more realistically in line with empirical observations. Regarding queue demand-based methods, Huntsinger and Rouphail (2011) and Moses et al. (2013) proposed queue demand-based link cost functions in which the demand at congestion is measured considering the queue. Building on Newell (1982), Cheng et al. (2022) and Zhou et al. (2022) developed a novel link cost function corresponding to spatial queue model with cubic arrival rate. Bliemer et al. (2014) and Bliemer and Raadsen (2020) proposed a non-separable path travel time function that models queuing delay by incorporating the node model into a static traffic assignment problem.

Unlike physical representations of traffic flow, these link cost functions may allow "flow" to exceed physical capacity. In traffic assignment models, the "flow" is not strictly related to physically measured flows but is more akin to demand, particularly under congested conditions (Kucharski and Drabicki, 2017; Cheng et al., 2022). This discrepancy complicates comparisons between assigned results (i.e., demand) from traffic assignment models under congested conditions and measured flow from traffic detectors, which will never exceed capacity.

To improve static traffic assignment models, various researchers have attempted to include residual queue considerations into the model (Lam and Zhang, 2000; Small and Verhoef, 2007; Huntsinger and Rouphail, 2011; Bliemer et al., 2014; Bliemer and Raadsen, 2020). This has proven to be a challenge as the residual queue stands for in general a characteristic of dynamic nature of traffic (Newell, 1968a, b, c). In addressing these issues, Lam and Zhang (2000) incorporated the residual queue concept in their model by proposing a capacity constrained traffic assignment model. Complementing this line of research, a seminal study by Bliemer et al. (2014) developed a quasi-dynamic traffic assignment model, which is indeed a capacity constrained traffic assignment model with consideration of residual queue. A first order node model was adopted to determine capacity constrained traffic flow, residual queue, and path travel time. Furthermore, Smith et al. (2019) and Bliemer and Raadsen (2020) introduced a steady-state, quasi-dynamic traffic assignment model with explicit consideration of link-exit capacities, queueing delays and explicit bounds on queue storage capacities to incorporate spillback effects. Despite these advancements, these models, apart from Bliemer and Raadsen (2020), have struggled to precisely depict situations where equilibrium link flow falls below the link's physical capacity during congested conditions.

A few emerging studies have tried to address these scenarios under congested conditions to represent the congested link flow more precisely (Cheng et al., 2021; Wu et al., 2021; Zhou et al., 2022; Cheng et al., 2023). A new s-shaped three-parameters traffic flow model was proposed by Cheng et al. (2021) to better represent the relationships between fundamental variables under a wide range of possible densities. Cheng et al. (2023) developed two stochastic fundamental diagram models for analyzing traffic state variations and explaining the sources of stochasticity in the road capacity. Interestingly, Wu et al. (2021) attempted to systematically investigate how to represent volume-to-capacity ratio in the BPR function, especially for a congested network. Even though BPR function is widely used in traffic assignment problems owning to its differentiability and monotone, several studies have acknowledged the limitations of BPR function. One of the major challenges is that the volume-to-capacity ratio under congested conditions (i.e., larger than one) cannot be empirically observed. Furthermore, the resultant link flows under congested conditions obtained from static traffic assignment models exceed the link physical capacity. It is challenging to avoid such an overestimation resulted from the utilization of BPR-type link cost function. Backward-bending link cost function consistent with the hydrodynamic traffic flow theory is one useful presentation of traffic. However, despite its good representation of traffic under congested conditions, the main difficulty to model the backward-bending link cost function into traffic assignment problems comes from its non-monotone property (Lo and Szeto, 2005).

The bottleneck model, introduced by Vickrey, was originally developed to address temporal distribution of commuters during peak periods (Vickrey, 1969). It has become one of the most commonly used link models for dynamic traffic assignment (DTA) problems (De Palma and Lindsey, 2002; Han et al., 2013a, b; Li et al., 2020). For example, De Palma and Lindsey (2002) used the Vickrey bottleneck model to address the DTA problem for evening commutes. Han et al. (2013a) and Han et al. (2013b) proposed a novel partial differential equation formulation of Vickrey's bottleneck model, which is particularly valuable in the context of analytical DTA. It is important to note that these earlier bottleneck models were designed to address the temporal distribution of traffic flow along a route within the DTA framework. In contrast, the model proposed in this paper focuses on an extended static traffic assignment model, which provides the spatial distribution of traffic flow across a road network during a specific time period.

Most of these models, as shown in Table 2, presented that achieving unique solutions for both link flows and residual queues remains a challenging endeavor. Unfortunately, this lack of solution uniqueness imposes limitations on the applicability of these models, even though more realistic. Therefore, it is interesting to develop a generalized link cost function to achieve precise representation of congested link flows and, importantly keeping the good mathematical properties of solutions (Shao et al., 2013; Shao et al., 2014; Shao et al., 2015; Ma and Qian, 2018; Fu et al., 2022).

In summary, while substantial research has been carried out in the field of traffic assignment, research gaps remain, particularly in the accurate representation of traffic flows under congested conditions, consideration of residual queue and queue-dependent capacity, and in the same time maintenance of good mathematical properties. The need for an innovative static traffic assignment model that accounts for these variables in a more realistic manner is evident.

## 2    Model Formulation

### 2.1   Notation and model assumptions

The variables and parameters used in this paper are listed in Table 3.

Table 3. Notation

| | |
|---|---|
| *Set* | |
| **A** | Set of links |
| **A̲** | Set of uncongested links |
| **Ā** | Set of congested links |
| $\mathbf{P}_{rs}$ | Set of paths linking the origin–destination (OD) pair *rs* |
| $\boldsymbol{\eta}_{ap}^{d}$ | Set of links downstream of link *a* along path *p* |
| $\boldsymbol{\eta}_{p}$ | Set of links along path *p* |
| **R, S** | Sets of origins and destinations, respectively |
| *Subscript* | |
| $a$ | Subscript representing link *a* |
| $p$ | Subscript representing path *p* |
| $rs$ | Subscript representing the OD pair *rs* |
| *Variable and parameter* | |
| $C_{a,max}$ | Physical capacity of link *a* (veh/hr) |
| $C_a$ | Capacity of link *a*, which is dependent on the residual queue (veh/hr) |
| $c_p$ | Travel cost on path *p*, including both the travel time and queuing delay on this path |
| $D_{rs}$ | Traffic demand from origin *r* to destination *s* |
| $f_p$ | Flow on path *p* |
| $\tilde{f}_p$ | Oversaturated path flow, including both the flow passing through the network during the considered period and the residual queue |
| $Q_a$ | Residual queue on link *a* during the considered period (veh/hr) |
| $Q_{ap}$ | Residual queue on link *a* caused by the flow from path *p* |
| $t_a$ | Travel time on link *a* (hr) |
| $t_{f,a}$ | Free flow travel time on link *a* (hr) |
| $u_{a,c}$ | Speed at the link physical capacity according to the link-based fundamental diagram of link *a* |
| $u_a$ | Speed on link *a* |
| $v_a$ | Flow on link *a* that passes through the link during the considered period (veh/hr) |
| $w_{rs}$ | Lagrange multipliers of the path flow conservation constraint |
| $\delta_{ap}$ | Link–path incidence, a binary variable representing whether link *a* is traversed by path *p* |
| $\mu_a$ | Lagrange multipliers of the link flow–queue relationship |
| $\alpha, \beta, m, n$ | Parameters of the proposed link cost function |
| $\gamma$ | Parameter of the link flow–queue relationship |
| $\varphi$ | Parameter relating to the queue-dependent link capacity |

The following assumptions are adopted to facilitate the presentation of the essential idea of this study:

(A1) The residual queue is treated as a (vertical) point queue without taking into account spillback effects in this paper (Lam and Zhang, 2000; Zhang et al., 2013; Bliemer et al., 2014; Li and Zhang, 2015). It is explicitly considered to determine the equilibrium state of a link during the concerned period (Bliemer et al., 2014; Lam and Zhang, 2000). The residual queue is assumed to be negatively related to the corresponding link flow under congested conditions.

(A2) The travel time on each link is a function of the link flow and residual queue on that link. The total travel time on a link consists of the travel time along the link and the queuing delay under congested conditions (Bliemer et al., 2014, 2020; Lam and Zhang, 2000). Under congested conditions, the queuing delay increases as the residual queue increases and the queue-dependent capacity decreases.

(A3) The study horizon is within one period. The carryover effects of a traffic queue across different periods are not considered in this paper (Lam and Zhang, 2000; Bliemer and Raadsen, 2020).

## 2.2 A generalized link cost function

In practice, the link capacity is influenced by the residual queue on that link. Under congested conditions, as a traffic queue forms, the vehicles in the queue occupy some road space at the link. These queued vehicles reduce the number of vehicles that can pass this link (i.e., the link capacity).

A residual queue is defined as the rate of vehicle queuing over the modeling period, which equals the number of vehicles retained on a link after the considered period (e.g., 08:00–09:00 am) divided by its duration. For a fixed number of vehicles entering the link, an increase in the number of vehicles that pass through the link leads to a decrease of the residual queue on that link. In subsequent sections of this paper, the term "link flow" refers to "traffic flow passing through the link." The relationship between the residual queue ($Q_a$ veh/hr) and link flow ($v_a$ in veh/hr) can be described mathematically as follows:

$$Q_a = \begin{cases} 0, & \text{if } u_a \geq u_{a,c} \\ \Gamma(v_a), & \text{if } u_a < u_{a,c} \end{cases} \quad \text{where } \frac{d\Gamma(v_a)}{dv_a} < 0, 0 \leq \Gamma(v_a) \leq M, \forall a \in \mathbf{A} \tag{1}$$

In Equation (1), $\Gamma(v_a)$ is a monotone decreasing polynomial function ranging from 0 to M. $u_a$ and $u_{a,c}$ are respectively the speed on link $a$ and the speed at link physical capacity according to the link-based fundamental diagram of link a. According to Equation (1), there is no residual queue under uncongested traffic conditions (i.e., $u_a \geq u_{a,c}$). However, under congested traffic conditions (i.e., $u_a < u_{a,c}$), the residual queue is negatively related to the link flow. For instance, the following link flow-queue relationship can be adopted:

$$Q_a = \begin{cases} 0, & \text{if } u_a \geq u_{a,c} \\ \frac{1}{\gamma}(C_{a,max} - v_a), & \text{if } u_a < u_{a,c} \end{cases}, \quad \forall a \in \mathbf{A} \tag{2}$$

where $\gamma$ is a parameter representing the extend that the residual queue affects the physical capacity. To govern the relationship between the link flow ($v_a$) and residual queue ($Q_a$) described in Equation (1), the following equation is used:

$$Q_a(\Gamma(v_a) - Q_a) = 0, \forall a \in \mathbf{A} \tag{3}$$

Equation (3) holds for either uncongested (i.e., $Q_a = 0$) or congested conditions (i.e., $\Gamma(v_a) - Q_a = 0$). The below-capacity link flow is implicitly managed through the relationship between flow and queue in Equations (1) – (3). As the residual queue should be non-negative ($Q_a \geq 0$), the link flow is constrained by the inequality in Equation (2) (i.e., $\frac{1}{\gamma}(C_{a,max} - v_a) \geq 0$) such that the inequality, $v_a \leq C_{a,max}$, holds.

The concept of the residual queue is further used to define the queue-dependent link capacity. In practice, the capacity of a link is defined as the number of vehicles that can pass through the link during a specific period. However, under congested conditions, the link capacity is not constant and can be reduced by a residual queue. This reduced link capacity should be related to the residual queue or traffic flow that is able to pass through the link. Therefore, the following representation of the queue-dependent capacity is proposed:

$$C_a(Q_a) = \begin{cases} C_{a,max}, & if\ Q_a = 0 \\ C_{a,max} - \gamma Q_a, & if\ Q_a > 0 \end{cases}, \forall a \in \mathbf{A} \tag{4}$$

However, the traditional link cost function (e.g., the BPR function) does not explicitly consider the residual queue and thus cannot address the queue-dependent capacity and corresponding queuing delay. The capacity used in the BPR function is typically the fixed physical or ultimate capacity (HCM, 2010; Wu et al., 2021). However, in this paper, we use queue-dependent capacity to account for the effects of traffic queues on link flow. Specifically, Physical/practical capacity is defined as the maximum flow rate that pass a given point on a link during a specified period without the oversaturated conditions, which is fixed and determined by the road attributes such as road type, carriageway width, frontage type (Transport Department, 2023). Ultimate capacity is defined as the maximum hourly flow rate when the road is under level of service (LOS) E (HCM, 2010; Wu et al., 2021; Zhou et al., 2022). Note that a range of LOSs from A to F was developed to quantify congestion: LOS A represents the free-flow state and LOS F represents breakdown. However, the queue-dependent capacity, referred to as the link (exit) capacity in this paper, is defined as the actual maximum hourly outflow rate under various traffic conditions, which is assumed to be influenced by the traffic queue on the link.

The main differences between physical or ultimate capacity and queue-dependent capacity are: 1) Physical or ultimate capacity is fixed and determined based on road attributes, while 2) Queue-dependent capacity is variable and influenced by traffic queues. The link capacity is calculated using a combination of Equations (1) and (4) in this paper. It should be noted that the capacity used in this paper functions more like an ultimate hourly capacity rather than a period-specific capacity. For period capacity, the peak period is considered as time windows lasting for several hours where travel demands are pre-specified (Cheng et al., 2023). An hour-to-period factor should be estimated to convert ultimate hourly capacity to period capacity (Wu et al., 2021; Zhou et al., 2022). The outstanding feature of the adopted capacity is its sensitivity to traffic queue under congested conditions. By proposing an hour-to-period factor, the adopted queue-dependent capacity can also be converted to queue-dependent period capacity.

To address these issues, a generalized link cost function (Equation 5) based on the traditional BPR function is proposed by introducing a queuing delay term.

$$t_a(v_a, Q_a) = t_{f,a}\left(1 + \beta\left(\frac{v_a}{C_a(Q_a)}\right)^n\right) + \alpha\left(\frac{Q_a}{C_a(Q_a)}\right)^m, \forall a \in \mathbf{A} \tag{5}$$

where $t_{f,a}$ is the free-flow travel time on link $a$. The first term, $t_{f,a}\left(1 + \beta\left(\frac{v_a}{C_a(Q_a)}\right)^n\right)$, is the traditional BPR function but includes the queue-dependent capacity, $C_a(Q_a)$. The second term, $\alpha\left(\frac{Q_a}{C_a(Q_a)}\right)^m$, represents the persistent queuing delay, defined as the residual queue ($Q_a$) divided by the corresponding queue-dependent capacity (Belezamo, 2020; Cheng et al., 2022). It is similar to the well-known queue clearance time in queuing theory, which is calculated in this paper by dividing the residual queue by the corresponding capacity (Akçelik and Rouphail, 1993; Lam and Zhang, 2000; Bliemer et al., 2014). Due to that the queuing delay even at the same flow level may be different for different road types and traffic conditions, the parameters $\alpha$ and $m$ that should be further calibrated are introduced in this term to decide the weight and shape of this queuing delay term (Raadsen and Bliemer, 2019a; Pan et al., 2022; Zhou et al., 2022; Pan et al., 2023). For example, by setting $\alpha = 1/2$ and $m = 1$, the residual queue on link 3 (with capacity 573 veh/hr) shown in Figure 2 is 54 veh/hr, and the last vehicle entering the queue takes 0.094 hr to dissipate. The average queuing delay then becomes 0.047 hr.

Interested readers may refer to Huntsinger and Rouphail (2011) for some calibration methods. In general, the values of $\alpha$ and $m$ for an expressway should be smaller than those for an urban road; this is similar to $\beta$ and $n$ in the BPR function. These values indicate that a queuing delay on an expressway is less sensitive to a residual queue than a queuing delay on an urban road. The difference is due to the greater number of flow interruptions and traffic signals on urban roads, which increase the sensitivity of the queuing delay to the residual queue.

It should be noted that travelers may give more weight to queuing delay than travel delay. This travelers' preference can be reflected by adjusting the weights in the proposed generalized link cost function in Equation (5) and will influence their route choice behaviors especially in stochastic user equilibrium (SUE) model. However, the value of weights on queuing delay and travelling delay will not affect the model properties (equivalence, solution uniqueness, etc.). It is an interesting topic to further investigate how the travelers' preferences on queuing delay and travelling delay affect the equilibrium results in SUE models.

Figure 3 shows an example of the proposed generalized link cost function. Under uncongested conditions, i.e., $C_a(Q_a = 0) = C_{a,max}$, the generalized link cost function (Equation (5)) becomes $t_a = t_{f,a}\left(1 + \beta\left(\frac{v_a}{C_{a,max}}\right)^n\right)$, which is exactly the BPR function as shown in Figure 3(a). Under congested conditions, a residual queue is present (i.e., $Q_a > 0$); according to Equation (4), $C_a(Q_a) = v_a$. From Equations (1) and (4), the generalized link cost function (Equation 5) under congested conditions becomes $t_a = t_{f,a}(1 + \beta) + \alpha(\frac{Q_a}{\Gamma^{-1}(Q_a)})^m$ and is shown in Figure 3(b). In other words, the link travel time under congested conditions is dependent only on the residual queue because the travel time for vehicles traveling on the link before queue location is fixed at $t_{f,a}(1 + \beta)$.

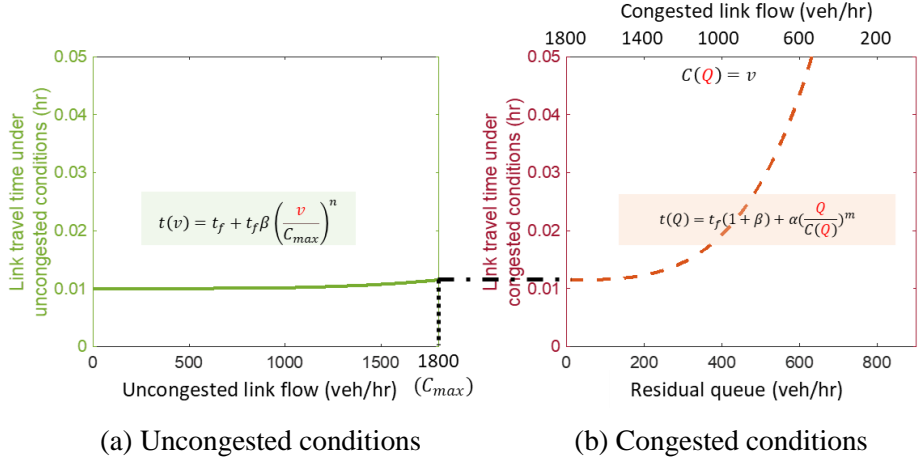

(a) Uncongested conditions     (b) Congested conditions

Figure 3. Example of the generalized link cost function

## 2.3 Static traffic assignment model with a residual queue and queue-dependent link capacity

To model the residual queue in a static traffic assignment model, the following path flow conservation equation is incorporated:

$$D_{rs} = \sum_{p \in P_{rs}} \left( f_p + \sum_{a \in \eta_p} Q_{ap} \right), \forall r \in \mathbf{R}, s \in \mathbf{S} \tag{6}$$

where $D_{rs}$ is the traffic demand from origin $r$ to destination $s$. $\mathbf{R}$ and $\mathbf{S}$ are sets of origins and destinations, respectively. $f_p$ is the path flow on path $p$. $Q_{ap}$ is the residual queue on link $a$ caused by the flow from path $p$. $\eta_p$ is the set of links along path $p$. By considering the residual queue, Equation (6) defines the traffic demand for any OD pair ($D_{rs}$) as equal to the sum of the total path flows that can complete the trip ($\sum_{p \in P_{rs}} f_p$) and the total queuing flow ($\sum_{p \in P_{rs}} \sum_{a \in \eta_p} Q_{ap}$) of all paths for that OD pair.

The vehicles in residual queue in a downstream link must also pass through the upstream links along the same path. Thus, the flow of any link $a$ on path $p$ ($v_{ap}$) should consist of two parts: 1) the path flows that can complete the trip in the concerned period, and 2) the flows that can pass through link $a$ but are retained as residual queues of the downstream link(s). Therefore, Equation (7) provides a definition of flow on a link:

$$v_a = \sum_{(rs)} \sum_{p \in P_{rs}} \delta_{ap} \left( f_p + \sum_{b \in \eta_{ap}^d} \delta_{bp} Q_{ap} \right), \forall a \in \mathbf{A} \tag{7}$$

where $\delta_{ap}$ is the link–path incidence, a binary variable that represents whether link $a$ is traversed by path $p$. $\eta_{ap}^d$ is the set of links downstream of link $a$ along path $p$. According to Equation (7), the flow on any link equals the sum of the path flows that can finish the trip ($\sum_{(rs)} \sum_{p \in P_{rs}} \delta_{ap} f_p$) and the flows that can pass through this link but are retained on its downstream links as a residual queue ($\sum_{(rs)} \sum_{p \in P_{rs}} \sum_{b \in \eta_{ap}^d} \delta_{ap} \delta_{bp} Q_{ap}$). The residual queue on a link equals the sum of the queues caused by all paths traversing that link and is defined in Equation (8).

$$Q_a = \sum_{(rs)} \sum_{p \in P_{rs}} Q_{ap}, \forall a \in \mathbf{A} \tag{8}$$

In addition, the following non-negativity condition should hold for all path flows:

$$f_p \geq 0, \ \forall p \in \mathbf{P}_{rs}, r \in \mathbf{R}, s \in \mathbf{S} \tag{9}$$

The proposed traffic assignment model that considers the queue-dependent capacity can be formulated as the following minimization problem:

$$\min J(\mathbf{v}, \mathbf{Q}) = \sum_{a \in \mathbf{A}} \left[ \int_0^{v_a} \left( t_{f,a} + t_{f,a} \beta \left( \frac{x_a}{C_a(Q_a)} \right)^n \right) dx_a + \int_0^{Q_a} \left( t_{f,a} + t_{f,a} \beta + \alpha \left( \frac{y_a}{C_a(y_a)} \right)^m \right) dy_a \right] \tag{10}$$

Subject to

Equations (1), (3), (4), and (6)–(9), (11)

where $\mathbf{v}$ is a vector of the link flow that can pass through the link ($v_a$), and $\mathbf{Q}$ is a vector of the residual queue ($Q_a$). The first part of the objective function (10) only computes for the links without queues. When a queue exists, $C_a(Q_a)$ becomes $v_a$ (link flow), making the first part of the objective function a fixed value. The second part of the objective function then computes for the links with queues.

Regarding the objective function of the proposed traffic assignment problem that considers the queue-dependent capacity (Equation 10), if there is no residual queue in a road network ($Q_a = 0, \forall a \in \mathbf{A}$), this objective is the same as that adopted in a traditional static traffic assignment problem. If a residual queue exists in a road network, the objective is to minimize the sum of the integrals of both travel time ($\int_0^{v_a} \left( t_{f,a} + t_{f,a} \beta \left( \frac{x_a}{C_a(Q_a)} \right)^n \right) dx_a$) on links without queue and queuing delay ($\int_0^{Q_a} \left( t_{f,a} + t_{f,a} \beta + \alpha \left( \frac{y_a}{C_a(y_a)} \right)^m \right) dy_a$) throughout the network.

## 3   Mathematical Properties

To better understand the properties of the proposed traffic assignment model that considers the queue-dependent capacity, the equivalency condition, smoothness of the objective function and uniqueness of the model are analyzed in this section.

It can be easily proved that the optimal solution of the proposed model always exists. Notice from constraints shown in Equations (1), (3), (4), and (6)–(9) that feasible regions of both decision variables (link flow $\mathbf{v}$ and residual queue $\mathbf{Q}$) are compact and nonempty sets. Also, the objective function in Equation (10) is continuous, though not smooth in the feasible regions. According to the Weierstrass extreme value theorem (Rusnock and Kerr-Lawson, 2005), the proposed model always has the optimal solution.

**Property 1** (Equivalency condition) The proposed minimization problem in Equations (10) and (11) is equivalent to the static traffic assignment model with queue-dependent capacity introduced in the previous section.

**Remark 1:** The additive and separable generalized cost is considered in this paper for simplicity. In other words, route travel cost $c_p$ can be calculated as the summation of the travel cost on links $c_a[= t_a(v_a, Q_a)]$ traversed by the route, which is mathematically expressed as $c_p = \sum_{a\in\eta_p} \delta_{ap} c_a \ \forall p \in \mathbf{P}_{rs}, r \in \mathbf{R}, s \in \mathbf{S}$. It should be noted that the path cost could be non-additive and non-separable if the number of vehicles in the queue is not stationary (Bliemer et al., 2014). Interested readers can refer to Bliemer et al. (2014) and Bliemer and Raadsen (2020) for detailed discussion.

The equilibrium conditions of the proposed model are defined following Sheffi (1985): a stable condition is reached only when no traveler can improve their generalized travel cost by unilaterally changing routes. To establish the equivalency between the optimization problem defined by Equations (10) and (11) and the proposed model, it is necessary to show that any flow patterns that solve the optimization problem also satisfy the equilibrium conditions. Given that the optimization problem [Equations (10) and (11)] is convex, as proved in Property 3, the Karush–Kuhn–Tucker (KKT) conditions are both necessary and sufficient for the global optimum. Therefore, this equivalency is demonstrated by proving that the KKT conditions for the optimization problem are identical to the equilibrium conditions.

Define an oversaturated path flow as $\tilde{f}_p = f_p + \sum_{a\in\eta_p} Q_{ap}$, which includes both the path flow passing through the network during the considered period and the residual queue. The KKT conditions of the minimization problem in Equations (10) and (11) are derived as follows:

$$\hat{f}_p \left( c_p + \sum_{a\in\eta_p} \delta_{ap}\mu_a - w_{rs} \right) = 0, \forall p \in \mathbf{P}_{rs}, r \in \mathbf{R}, s \in \mathbf{S} \tag{12a}$$

$$c_p + \sum_{a\in\eta_p} \delta_{ap}\mu_a - w_{rs} \geq 0, \forall p \in \mathbf{P}_{rs}, r \in \mathbf{R}, s \in \mathbf{S} \tag{12b}$$

$$D_{rs} = \sum_{p\in \mathbf{P}_{rs}} \hat{f}_p, \forall r \in \mathbf{R}, s \in \mathbf{S} \tag{12c}$$

where $c_p$ is the travel cost on path $p$, including both the traveling time on uncongested sections and queuing delay on this path. $w_{rs}$ and $\mu_a$ are the Lagrange multipliers of the path flow conservation constraint (Equation 6) and the link flow–queue relationship (Equation 3), respectively.

For a given path, the equivalency conditions hold for the following two possible combinations of path flow, path travel time and queuing delay:

(a) If the flow on the path is zero (i.e., $\hat{f}_p = 0$), the sum of the travel time and queuing delay on this path referred to as the generalized travel cost (i.e., $c_p + \sum_{a\in\eta_p} \delta_{ap}\mu_a$) must be greater than or equal to the OD-specific Lagrange multiplier ($w_{rs}$).

(b) If the flow on the path is positive (i.e., $\hat{f}_p > 0$), the Lagrange multiplier of the given OD pair equals the generalized travel cost on the paths connecting this pair (i.e., $c_p + \sum_{a\in\eta_p} \delta_{ap}\mu_a$).

The above conditions (a) and (b) mean that travelers only choose the paths with the lowest generalized travel cost (i.e., $c_p + \sum_{a\in\eta_p} \delta_{ap}\mu_a = w_{rs}$), and thus the flow on these paths is positive ($\hat{f}_p > 0$), whereas there is no flow ($\hat{f}_{p\prime} = 0$) on the paths with higher generalized travel costs (i.e., $c_{p\prime} + \sum_{a\in\eta_{p\prime}} \delta_{ap\prime}\mu_a > w_{rs}$). This result indicates that the KKT conditions of the proposed minimization problem are equivalent to the equilibrium conditions. The proof of equivalence is completed.

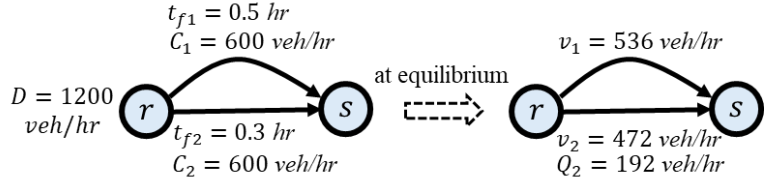

Figure 4. Illustration of model properties

In the two-link network shown in Figure 4, travelers departing from the origin can choose either the upper link (link 1) or the lower link (link 2), both of which have the same physical capacity. However, due to different free-flow travel times on these two links, the generalized link costs differ. While the network can handle 1200 veh/hr without queueing if 600 veh/hr is assigned to each link, this situation does not represent an equilibrium for individual travelers. At equilibrium, 536 veh/hr use the upper link and 472 veh/hr use the lower link, leaving a residual queue of 192 veh/hr on the lower link. This ensures that the generalized costs for the two links are equal. This example illustrates that, due to the longer distance of alternative routes, some travelers prefer to wait in the queue rather than taking the longer route.

**Remark 2:** One can easily see that the generalized link cost function in Equation (5) is not a smooth function, given the definition of queue-dependent capacity and the fact that this queue-dependent link capacity should be equal to the link physical capacity ($C_{a,max}$) under uncongested conditions. As a result, the objective function of the proposed traffic assignment model that considers the queue-dependent capacity (Equation 10) is also inherently not smooth.

To mathematically prove the objective function is not smooth, we need to show that it is not continuously differentiable. Notice from the definition of queue-dependent capacity, showing that the objective function is not continuously differentiable at the point $\{v, Q | v_{a'} = C_{a',max}, Q_{a'} = 0\}$ can achieve the proof. The proof is shown below:

The partial derivative of the objective function with respect to $v_{a'}$ on the point $\{v, Q | v_{a'} = C_{a',max}, Q_{a'} = 0\}$ is

$$\lim_{v_{a'} \to C_{a',max}} \frac{J(v, Q | Q_{a'} = 0) - J(v, Q | v_{a'} = C_{a',max}, Q_{a'} = 0)}{v_{a'} - C_{a',max}} =$$

$$\lim_{v_{a'} \to C_{a',max}} \frac{\int_0^{v_{a'}} \left(t_{f,a'} + t_{f,a'} \beta \left(\frac{x_a}{C_{a',max}}\right)^n\right) dx_a - \int_0^{C_{a',max}} \left(t_{f,a} + t_{f,a} \beta \left(\frac{x_a}{C_{a',max}}\right)^n\right) dx_a}{v_{a'} - C_{a',max}} \neq 0.$$

The partial derivative of the objective function with respect to $Q_{a'}$ on the point $\{v, Q | v_{a'} = C_{a',max}, Q_{a'} = 0\}$ is

$$\lim_{Q_{a'} \to 0} \frac{J(v, Q | v_{a'} = C_{a',max}) - J(v, Q | v_{a'} = C_{a',max}, Q_{a'} = 0)}{Q_{a'} - 0} =$$

$$\lim_{Q_{a'} \to 0} \frac{\int_0^{C_{a',max}} \left(t_{f,a} + t_{f,a} \beta \left(\frac{x_a}{C_{a',max}}\right)^n\right) dx_a - \int_0^{C_{a',max}} \left(t_{f,a} + t_{f,a} \beta \left(\frac{x_a}{C_{a',max}}\right)^n\right) dx_a}{Q_{a'} - 0} = 0.$$

Due to the partial derivative of the objective function at point $\{v, Q | v_{a'} = C_{a',max}, Q_{a'} = 0\}$ is not continuous, the objective function is not smooth. The proof is completed.

Note that the original formula in Equation (10) is a piece-wise nonlinear function, which leads to difficulties in doing integral and derivative, discovering model properties, and solving the problem. Even though some algorithms like sub-gradient method can be used to solve the non-smooth problem, smoothing the original formula will bring more advantages to analyze the proposed models and apply the proposed model to traffic management and planning. Therefore, we propose in this paper a sigmoid-function based reformulation to smooth the objective function in Equation (10) and the link cost function (Sheu and Pan, 2014; Du and Gong, 2016).

**Property 2** (Smoothness of the objective function) The objective of the minimization problem (10) can be smoothed by reformulating the objective function as follows:

$$\min J(\mathbf{v}, \mathbf{Q}) = \sum_{a \in \mathbf{A}} \left[ \int_0^{v_a} \left( t_{f,a} + t_{f,a} \beta \left( \frac{x_a}{C_{a,max}} \right)^{n\varphi^{-Q_a}} \right) dx_a + \int_0^{Q_a} \left( t_{f,a} + t_{f,a} \beta + \alpha \left( \frac{y_a}{C_a(y_a)} \right)^m \right) dy_a \right] \quad (13)$$

where $\varphi$ is a parameter related to the queue-dependent link capacity ($\varphi \geq 1$). Normally, $\varphi$ can be defined as Euler's number (i.e., $\varphi = e$), as in the sigmoid function. The first term in Equation (13) is the smoothed term by introducing the parameter $\varphi$ and eliminating the residual queue-dependent capacity, according to sigmoid function, while the second term with queue-dependent capacity is identical to its original objective function in Equation (10). The proof of the equivalency between original and smoothed objective functions shown respectively in Equations (10) and (13) is provided in Appendix C.

It can be mathematically demonstrated that the proposed model considering residual queue and queue-dependent capacity is more generalized than some traditional models. Specifically, if residual queue is excluded in the model (i.e., $Q_a = 0$), the proposed model becomes the traditional STA model, which does not guarantee that capacity is not exceeded as shown in Equation (14). Under specific conditions, the proposed model aligns with the capacity-constrained STA model as shown in Equation (15).

(a) For $Q_a = 0 \ \forall a \in \mathbf{A}$ (uncongested conditions), the queue-dependent capacity is equal to the link physical capacity ($C_{a,max}$). The generalized smooth objective function becomes:

$$\min J(\mathbf{v}, \mathbf{Q}) = \sum_{a \in \mathbf{A}} \left[ \int_0^{v_a} \left( t_{f,a} + t_{f,a} \beta \left( \frac{x_a}{C_{a,max}} \right)^n \right) dx_a \right] \quad (14)$$

The proposed model becomes the **traditional static traffic assignment** model without a capacity constraint.

(b) For $Q_a > 0 \ \exists a \in \mathbf{A}$ and $\varphi = 1$, the generalized smooth objective function becomes:

$$\min J(\mathbf{v}, \mathbf{Q}) = \sum_{a \in \mathbf{A}} \left[ \int_0^{v_a} \left( t_{f,a} + t_{f,a} \beta \left( \frac{x_a}{C_{a,max}} \right)^n \right) dx_a + \int_0^{Q_a} \left( t_{f,a} + t_{f,a} \beta + \alpha \left( \frac{y_a}{C_a(y_a)} \right)^m \right) dy_a \right] \quad (15)$$

The proposed model becomes the traffic assignment model with a fixed link physical capacity ($C_{a,max}$) and residual queue.

(c) For $Q_a > 0 \ \exists a \in \mathbf{A}$, $\varphi > 1$ and $\varphi \to +\infty$, the proposed model becomes the proposed traffic assignment model with a queue-dependent capacity. Specifically, as $Q_a > 0$ and $\varphi \to +\infty$, $\varphi^{-Q_a} \cong 0$ and $(x_a/C_{a,max})^{n\varphi^{-Q_a}} = 1$. This approach aims to replicate the scenario wherein the link capacity is equal to the link flow when there is a queue. Thus, under congested conditions, the objective function becomes:

$$\min J(\mathbf{v}, \mathbf{Q}) = \sum_{a \in \mathbf{A}} \left[ \int_0^{Q_a} \left( t_{f,a} + t_{f,a} \beta + \alpha \left( \frac{y_a}{C_a(y_a)} \right)^m \right) dy_a \right] \quad (16)$$

When $\varphi \to +\infty$, for links under uncongested conditions, the smoothed objective function aims to minimize the integral of the BPR function, and the link capacity is equal to the physical capacity (Equation 14). For links under congested conditions, however, the smoothed objective function aims to minimize the integral of the queuing delay (Equation 16), and the link capacity is equal to the link flow, which is expressed in terms of the residual queue. In other words, the reformulated smoothed objective function in Equation (13) is the same as that shown in Equation (10) with integration of the queue-dependent capacity constraints (Equations (3) and (4)) when $\varphi \to +\infty$. In this study, $\varphi$ is set as Euler's number ($e$) when approximating Equation (10) with a smooth function (Equation 13).

**Property 3** (Uniqueness of the proposed model) Given that the traffic conditions (uncongested or congested) on all links are known, the equivalent minimization problem of the proposed queue-dependent traffic assignment **has a unique solution** regarding the link flow and residual queue under the following conditions: (i) $\frac{\partial t(v_a, Q_a)}{\partial v_b} = 0$ and (ii) $\frac{\partial t(v_a, Q_a)}{\partial Q_b} = 0$.

Conditions (i) and (ii) mean that the total travel time on link $a$ is independent of the link flow and residual queue on other links. These conditions also reflect the ignorance of the queue spillback effect in this study.

To prove the uniqueness of the proposed model, an equivalent model is formulated by introducing a separable travel time function as follows. The link set is separated into uncongested and congested link sets $\underline{\mathbf{A}}$ and $\overline{\mathbf{A}}$, respectively, with $\underline{\mathbf{A}} \cup \overline{\mathbf{A}} = \mathbf{A}$. There is no residual queue at equilibrium in the uncongested link set, whereas there exists a residual queue on links in the congested link set. Therefore, the objective function in Equation (10) can be reformulated as

$$\min J(\mathbf{v}, \mathbf{Q}) = \sum_{a \in \underline{\mathbf{A}}} \left[ \int_0^{v_a} \left( t_{f,a} + t_{f,a} \beta \left( \frac{x_a}{C_{a,max}} \right)^n \right) dx_a \right] +$$

$$\sum_{a \in \overline{\mathbf{A}}} \left[ \int_0^{Q_a} \left( t_{f,a} + t_{f,a} \beta + \alpha \left( \frac{y_a}{C_a(y_a)} \right)^m \right) dy_a \right] \quad (17)$$

Here, a link should be either uncongested or congested. If link $a$ is uncongested, the link flow on this link is the only decision variable. If link $a$ is congested, the residual queue on this link becomes the only decision variable in Equation (17). The following proof demonstrates that the objective function is strictly convex with respect to the link flow and residual queue. Note that the objective function in Equation (17) consists of two separable terms, one for uncongested links and one for congested links. Each term of the objective function is smooth. To prove the convexity of the original objective function, we separately prove the convexity of its two terms by computing their Hessian matrices respectively below.

First, the first-order derivative of the objective function is derived. With respect to the link flow on link $a$, which is uncongested, the first-order derivative should be

$$\frac{\partial J(\mathbf{v}, \mathbf{Q})}{\partial v_a} = t_{f,a} + t_{f,a} \beta \left( \frac{v_a}{C_{a,max}} \right)^n \quad (18a)$$

With respect to the link residual queue on link $b$, which is congested, the first-order derivative should be

$$\frac{\partial J(\mathbf{v}, \mathbf{Q})}{\partial Q_b} = t_{f,b} + t_{f,b}\beta + \alpha \left(\frac{Q_b}{C_b(Q_b)}\right)^m \tag{18b}$$

Second, the second-order derivative of the objective function for the uncongested links in Equation (18c) can be derived as follows:

$$\frac{\partial^2 J(\mathbf{v}, \mathbf{Q})}{\partial^2 v_a} = \frac{nt_{f,a}\beta}{C_{a,max}^n} v_a^{n-1} > 0 \tag{18c}$$

Also, considering that $\frac{d\Gamma(v_a)}{dv_a} < 0$ defined in Equation (1), the second-order derivative of the objective function for the congested links in Equation (18d) can be derived as:

$$\frac{\partial^2 J(\mathbf{v}, \mathbf{Q})}{\partial^2 Q_b} = \alpha m \left(\frac{Q_b}{C_b(Q_b)}\right)^m \frac{C_b(Q_b) - Q_b(dv_b/d\Gamma(v_b))}{[C_b(Q_b)]^2} > 0 \tag{18d}$$

From the first-order derivatives presented in Equations (18a) and (18b) and the assumption that the total travel time on link $a$ is independent of the link flows and residual queues on other links, we obtain

$$\frac{\partial^2 J(\mathbf{v}, \mathbf{Q})}{\partial v_a \partial Q_b} = \frac{\partial^2 J(\mathbf{v}, \mathbf{Q})}{\partial Q_b \partial v_a} = 0 \tag{18e}$$

Therefore, the off-diagonal elements of the Hessian are all zero and all of the diagonal elements are positive, thus ensuring a positive-definite Hessian matrix. The objective is thus strictly convex with respect to the link flow and residual queue. As the feasible region (Equations (6)–(9)) is also convex, the proposed traffic assignment model has a unique solution.

It has been proved that the proposed traffic assignment model has a unique solution with respect to link flow and residual queue. However, the path flow could not be unique. An illustrating example is shown in Figure 5 below to explain the non-unique path flow. Assume that the resultant unique link flow and residual queue ($\mathbf{v}$, $\mathbf{Q}$) are obtained and presented near the corresponding links in Figure 5. The solution of path flows ($f_1, f_2, f_3, f_4$) can be either (25,10,65,50) or (15,20,75, 40), and both satisfy the flow conservations shown in Equations (6) – (9). Therefore, the path flow solutions under some cases for the proposed model may not be unique.

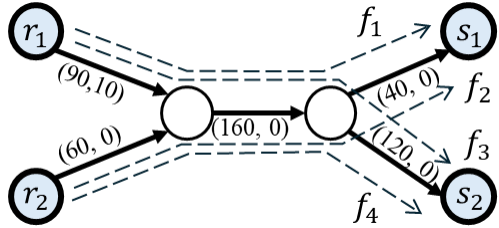

Figure 5. Illustrating network with non-unique path flow solutions

## 4    Solution Algorithm

The consideration of capacity constraint when solving a traffic assignment model is commonly recognized to increase the difficulty of the model (Nie et al., 2004; Li and Chen, 2023). The solution algorithm for a traffic assignment model that considers the capacity constraint may encounter convergence issues, resulting in infeasible or non-optimal solutions. Reformulation of the objective function (Equation (13)) with the integration of queue-dependent capacity is advantageous because this makes it unnecessary to consider the queue-dependent capacity in the constraints. This reformulation of the proposed model benefits the solution algorithm and allows the proposed static traffic assignment model with a queue-dependent capacity to be solved efficiently.

Including the definition of an oversaturated path flow ($\tilde{f}_p$), Equation (6) becomes

$$D_{rs} = \sum_{p \in P_{rs}} \tilde{f}_p, \forall r \in R, s \in S \tag{19}$$

Define $\eta_{ap}^u$ as the set of links upstream of link $a$ along path $p$. Equation (7) becomes

$$v_a = \sum_{(rs)} \sum_{p \in P_{rs}} \delta_{ap} \left( \tilde{f}_p - Q_{ap} - \sum_{b \in \eta_{ap}^u} \delta_{bp} Q_{ap} \right), \forall a \in A \tag{20}$$

For simplicity, define $Q'_a = \sum_{(rs)} \sum_{p \in P_{rs}} \sum_{b \in \eta_{ap}^u} \delta_{ap} \delta_{bp} Q_{ap}$ as the total residual queue on the upstream links of link $a$. Then, Equation (20) becomes

$$v_a = \sum_{(rs)} \sum_{p \in P_{rs}} \delta_{ap} \tilde{f}_p - Q_a - Q'_a, \forall a \in A \tag{21}$$

Drawing on Equation (21), and regarding the oversaturated path flow and residual queue as the decision variables instead of the link flow and residual queue, the objective of the minimization problem can be expressed as

$$\min J(\tilde{\mathbf{f}}_p, \mathbf{Q}) = \sum_{a \in A} \int_0^{\sum_{(r,s)} \sum_{p \in P_{rs}} \delta_{ap} \tilde{f}_p - q_a - q'_a} \left( t_{f,a} + t_{f,a} \beta \left( \frac{x_a}{C_{a,max}} \right)^{n\varphi^{-Q_a}} \right) dx_a +$$

$$\sum_{a \in A} \int_0^{Q_a} \left( t_{f,a} + t_{f,a} \beta + \alpha \left( \frac{y_a}{\Gamma^{-1}(y_a)} \right)^m \right) dy_a \tag{22}$$

The vicinity of the oversaturated path flow is defined as $\Omega_f = \{D_{rs} = \sum_{p \in P_{rs}} \tilde{f}_p \text{ and } \tilde{f}_p \geq 0, \forall p \in P_{rs}, r \in R, s \in S\}$, and the vicinity of the residual queue is defined as $\Omega_Q = \{Q_a \geq 0 \text{ and } \Gamma^{-1}(Q_a) \geq 0, \forall a \in A\}$.

To simplify the problem for implementation of an efficient solution algorithm, the idea of an alternating minimization algorithm is adopted to solve subproblems with different decision variables ($\tilde{\mathbf{f}}_p$ and $\mathbf{Q}$) (Tseng, 1991). Therefore, a GP-AM algorithm (i.e., gradient projection – alternating minimization) is proposed to solve the above problem as follows.

**Step 0** Initialization: $k = 0$, $\tilde{\mathbf{f}}_p^0 = 0$, $\mathbf{Q}^0 = 0$.

**Step 1** Solve for the oversaturated path flow

$$\tilde{\mathbf{f}}_p^{k+1} = \arg\min_{\tilde{\mathbf{f}}_p \in F} J(\tilde{\mathbf{f}}_p, \mathbf{Q}^k) \tag{23a}$$

Implement the conventional GP (i.e., gradient projection) algorithm (LeBlanc et al., 1975) to solve this subproblem in Equation (23a).

**Step 2** Solve for the residual queue

$$\mathbf{Q}^{k+1} = \arg\min_{\tilde{\mathbf{f}}_p \in F} J(\tilde{\mathbf{f}}_p^{k+1}, \mathbf{Q}) \tag{23b}$$

Implement a modified GP algorithm to solve this subproblem in Equation (23b). The direction of descent in this subproblem is the partial derivative of the objective function in Equation (22) with respect to the residual queue (**Q**).

**Step 3** Check the convergence as: $max\left(\left\|\tilde{\mathbf{f}}_p^{k+1} - \tilde{\mathbf{f}}_p^k\right\|_\infty, \left\|\mathbf{Q}^{k+1} - \mathbf{Q}^k\right\|_\infty\right) \leq \epsilon$.

If the above stopping criteria are satisfied, stop; otherwise, $k := k + 1$. Return to step 1.

$\|\mathbf{x}\|_\infty = \max_i |x_i|$, which gives the largest magnitude of each element of a vector. After implementing the above GP-AM algorithm, the oversaturated path flow and residual queue can be obtained. Using an approach based on Equation (21), the link flow for all links in the road network can then be obtained.

Following Equation (17), the objective function can be written as $\min J(\mathbf{v}, \mathbf{Q}) = J_1(\mathbf{v}) + J_2(\mathbf{Q})$, The functions $J_1: \mathbb{R}^{|A|} \to [0, \infty]$ and $J_2: \mathbb{R}^{|A|} \to [0, \infty]$ are closed and proper convex functions, and subdifferentiable over their domain. Note that for any $\tilde{\mathbf{f}}_p \in \Omega_f$, $\tilde{\mathbf{Q}} \in \Omega_q$, the problems in Equations (23a, b) have minimizers (LeBlanc et al., 1975). We will show that the limit points of the sequence generated by the alternating minimization method are stationary points of the problem.

**Lemma 1**. For $k \geq 0$, define $\mathbf{x}^k := (\tilde{\mathbf{f}}_p^k, \mathbf{Q}^k)$, $\mathbf{x}^{k+1/2} := (\tilde{\mathbf{f}}_p^{k+1}, \mathbf{Q}^k)$, $J^k := J(\mathbf{x}^k)$, $J^{k+1/2} := J(\mathbf{x}^{k+1/2})$. Let $\{\mathbf{x}^k\}_{k \geq 0}$ be the sequence generated by the alternating minimization method. Then for any $k \geq 0$, the following inequality holds (Beck, 2015):

$$J(\mathbf{x}^k) - J\left(\mathbf{x}^{k+\frac{1}{2}}\right) \geq \frac{1}{2L_1}\left\|G_{L_1}^1(\mathbf{x}^k)\right\|^2 \tag{24}$$

where $G_{L_1}^1$ is the partial gradient mapping for any $L_1 > 0$ (Nesterov, 2018).

**Theorem 1** (rate of convergence of partial gradient mapping). Let $\{\mathbf{x}^k\}_{k \geq 0}$ be the sequence generated by the alternating minimization method. Then for any $n \geq 0$, we have (Beck, 2015):

$$\min_{k=0,1,2,\ldots,n}\left\|G_{L_1}^1(\mathbf{x}^k)\right\| \leq \frac{\sqrt{2L_1(H(x_0) - H^*)}}{\sqrt{n+1}} \tag{25}$$

The proof of both Lemma and Theorem can be found in Beck (2015).

## 5   Numerical examples

To demonstrate the characteristics and effectiveness of the proposed model, numerical examples are applied to a hypothetical small road network and a large real-world road network. In this hypothetical road network, experiments are conducted to (1) determine the effects of the queue-dependent link capacity on the traffic assignment results, (2) determine the effects of the residual queue on the traffic assignment results and (3) conduct a sensitivity analysis of the parameters in the link flow–queue relationship. In the large real-world road network, the applicability and convergence of the proposed solution algorithm are verified.

The small road network consists of six nodes and seven links, as shown in Figure 6. In this road network, there are two OD pairs, Node 1–Node 3 and Node 2–Node 4, and four paths. The path information is depicted in Figure 6. Table 4 presents the link attributes in this network. In the following numerical examples, the link flow-queue relationship in Equation (2) is adopted, and values of $\gamma$ and $\omega$ are set to be 0.5 and 1, respectively. Subject to the results of the empirical validation on these values, the model formulation and properties will be reviewed accordingly.

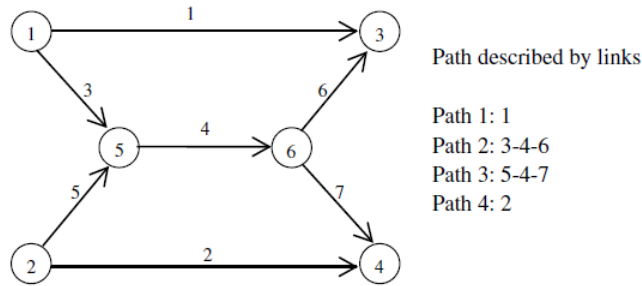

Figure 6. A hypothetical small road network

Table 4. Link attributes in the hypothetical small road network

| Attribute | Link No. | | | | | | |
|---|---|---|---|---|---|---|---|
| | 1 | 2 | 3 | 4 | 5 | 6 | 7 |
| $t_f$ (hr) | 0.417 | 0.417 | 0.167 | 0.167 | 0.167 | 0.133 | 0.133 |
| $u_f$ (km/hr) | 70 | 70 | 70 | 70 | 70 | 70 | 70 |
| $c_{max}$ (veh/hr) | 1800 | 1800 | 1800 | 2400 | 1800 | 1500 | 1500 |
| $l$ (km) | 29.17 | 29.17 | 11.67 | 11.67 | 11.67 | 9.33 | 9.33 |

The results of the proposed traffic assignment model on all links are presented in Table 5. The generalized cost for the network is 2.206 hr, and the resultant path travel times for different paths at equilibrium are uniformly 0.614 hr. As shown in Table 5, link 4 serves as the bottleneck link, with residual queue forming only on this link.

Table 5. Assignment results on all links

| Link No. | 1 | 2 | 3 | 4 | 5 | 6 | 7 |
|---|---|---|---|---|---|---|---|
| Link flow (veh/hr) | 1775 | 1775 | 1225 | **2350** | 1225 | 1775 | 1775 |
| Residual queue (veh/hr) | 0 | 0 | 0 | **100** | 0 | 0 | 0 |
| Travelling time (hr) | 0.614 | 0.614 | 0.185 | 0.271 | 0.185 | 0.158 | 0.158 |
| Queuing delay (hr) | 0 | 0 | 0 | **0.021** | 0 | 0 | 0 |
| Generalized link cost (hr) | 0.614 | 0.614 | 0.185 | 0.292 | 0.185 | 0.158 | 0.158 |

## 5.1 Effects of queue-dependent link capacity

Two scenarios with fixed (Nie et al., 2004) and queue-dependent (proposed model) link capacity constraints are adopted to demonstrate the effects of the queue-dependent link capacity on the traffic assignment results. The OD demands of OD pairs 1–3 and 2–4 are both set as 3000 veh/h. The results on Link 4, the bottleneck, are presented in Table 6. In the model with a fixed capacity constraint, the link physical capacity is used in the capacity constraint. Table 6 indicates that the link flow on Link 4 estimated by the traffic assignment model with a fixed capacity equals the link physical capacity of Link 4 (2400 veh/hr). However, in the model with a queue-dependent link capacity constraint, the link capacity is reduced to 2350 veh/hr when a residual queue forms. As a result, the link flow of Link 4 under equilibrium becomes 2350 veh/hr, which is less than the link physical capacity and should be consistent with the link flow observed by the traffic detector under congested conditions (Figure 1).

The model with a fixed capacity yields a residual queue of 68 vehicles on link 4, in contrast to the queue of 100 vehicles yielded by the model with a queue-dependent capacity. This difference, which exceeds 30%, is due to the consideration of a queue-dependent capacity (Equation (3)) in the proposed model. Furthermore, the increase in residual queue leads to an approximately 30% larger queuing delay for the model with a queue-dependent capacity (0.021 hr) than for the model with a fixed capacity (0.014 hr). The results in Table 6 show that the proposed model with a queue-dependent link capacity can generate link flows that are less than the link physical capacity under congested conditions. This result is consistent with observations from traffic detectors. It can be also observed that links 3 and 5 do not retain residual queue. This could be because that these links are each traversed by only one path (i.e., path 2 and path 3, respectively), and the spillback effect is not considered.

Table 6. Assignment results of models with fixed and queue-dependent link capacity constraints

| Model | Link flow (veh/hr) | Link capacity (veh/hr) | Residual queue (veh/hr) | Queuing delay (hr) | Generalized link cost (hr) |
|---|---|---|---|---|---|
| | | | Link 3/5 | | |
| Fixed capacity (Physical capacity) constraint | 1234 | 1800 | 0 | 0 | 0.185 |
| Queue-dependent capacity constraint | 1225 | 1800 | 0 | 0 | 0.185 |
| | | | Link 4 | | |
| Fixed capacity (Physical capacity) constraint | 2400 | 2400 | 68 | 0.014 | 0.264 |
| Queue-dependent capacity constraint | 2350 | 2350 | 100 | 0.021 | 0.271 |

## 5.2  Effects of residual queue

The proposed model is able to address the inconsistencies between the estimated link flows from the traffic assignment model and observed link flows under congested conditions by introducing the residual queue, the queue-dependent capacity and an extra queuing delay term. With these additional features, the estimated link flow under congested conditions can decrease as the residual queue increases in the scenario of continuously increasing traffic demand. However, the link flow increases with increasing traffic demands under uncongested conditions. To demonstrate these effects of a residual queue on the assignment results, the traffic demand for OD pair 1–3 is set to vary from 1000 to 6000 veh/hr, while the traffic demand for OD pair 2–4 is fixed at zero. The resultant link flow and residual queue on link 3 are examined as shown in Figure 7.

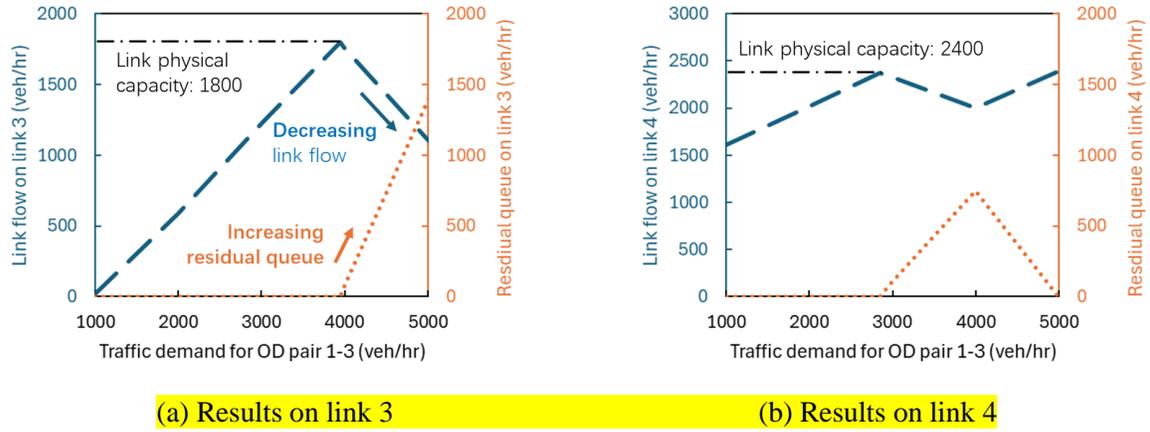

(a) Results on link 3              (b) Results on link 4

Figure 7. Model results of the link flow and residual queue under various demands

Figure 7 shows that as OD demand for OD pair 1-3 increases, the link flow on both links 3 and 4 begins to decrease once it reaches their physical capacity; at which point the residual queue starts to form. Interestingly, Figure 7(b) shows that, the residual queue on Link 4 does not continuously increase with higher demand; instead, it decreases when the demand reaches about 4000 veh/hr. This phenomenon may be due to the upstream effect from Link 3, where vehicles start to be queue when the demand reaches about 4000 veh/hr, reducing the flow into link 4 and consequently decreasing the residual queue on Link 4.

Figure 7 demonstrates the superiority of the proposed model such that when the traffic demands increase, the link flow under equilibrium initially increases but then decreases after reaching the link physical capacity. Meanwhile, a residual queue begins to form when the link flow reaches the physical capacity. This assignment result is more practical than those yielded by traditional models and coincides with the data collected from traffic detectors.

### 5.3 Sensitivity analysis on the key parameter

Furthermore, the effect of a residual queue on the corresponding link flow is affected by the choice of parameter $\gamma$ as described in Equation (1). The value of $\gamma$ is normally within (0,1). Basically, for a given link flow, the larger the value of $\gamma$ (i.e., close to 1), the smaller the residual queue. The following sensitivity analysis is conducted to show the effect of the choice of $\gamma$ on the assignment results.

Table 7. Sensitivity analysis of the effects of parameter $\gamma$ in the link flow–queue relationship

| $\gamma$ | Link flow (veh/hr) | Residual queue (veh/hr) | Queuing delay (hr) | Traveling time (hr) |
|---|---|---|---|---|
| 0 | 2400 | 176 | 0.037 | 0.287 |
| 0.2 | 2358 | 212 | 0.045 | 0.295 |
| 0.5 | 2250 | 301 | 0.067 | 0.317 |
| 0.7 | 2114 | 409 | 0.097 | 0.347 |
| 0.9 | 1861 | 599 | 0.161 | 0.411 |

Interestingly, the data in Table 7 show that as $\gamma$ increases, the residual queue increases while the link flow under equilibrium decreases. As $\gamma$ increases, the queuing delay increases such that the link travel time increases. Thus, the link flow and residual queue on this link are reduced, as less traffic is assigned to this link. According to Equation (2), as the link flow decreases ($v_a$), the residual queue ($q_a$) increases. Such an increase should outweigh the decrease in the residual queue caused by the increase in $\gamma$ and thus lead to an increase in the residual queue (Table 7).

In practice, $\gamma$ may be link-dependent and could be affected by the overall traffic conditions at a link, such as the traffic flow mix (i.e., the proportions of various types of vehicles) and the proportion of green time on the signalized intersection allocated to that link. For example, as a large number of heavy vehicles on a road may increase the residual queue, an increase in the number of heavy vehicles on a road should lead to an increase in $\gamma$ according to the above-described sensitivity analysis. Furthermore, as the proportion of green time of the signalized intersection allocated to a given link decreases, the number of vehicles that can pass the link (smaller link flows) also decreases and, thus, $\gamma$ increases. In practice, $\gamma$ can be simply calibrated using observation data from video-based traffic detectors when the link flow and residual queue are available.

### 5.4 Performance on a large real-world road network

To further illustrate the performance of the proposed model and solution algorithm, we investigated a large real-world network in Melbourne, Australia with general modeling network specification (Nourmohammadi et al., 2021; Lu and Zhou, 2023). This network features a grid pattern with 2,077 nodes and 4,223 links, as shown in Figure 8. For our study, the path set is predefined and fixed. The width of the green lines and red lines in the figure indicates the amount of assigned link flow and residual queue, respectively. Approximately 10% of the links (418 links) retained a residual queue during the study period. The results shown in Figure 8 highlight potential links, particularly those heavily congested ones marked in red, that could benefit from improvements or the implementation of restriction policies (e.g., congestion charging) to enhance network performance further.

The convergence of our solution algorithm is visually presented in Figure 9, which details the progression across different iterations. It should be noted that it took 756 seconds for the proposed GP-AM algorithm to converge at the 138th iteration.

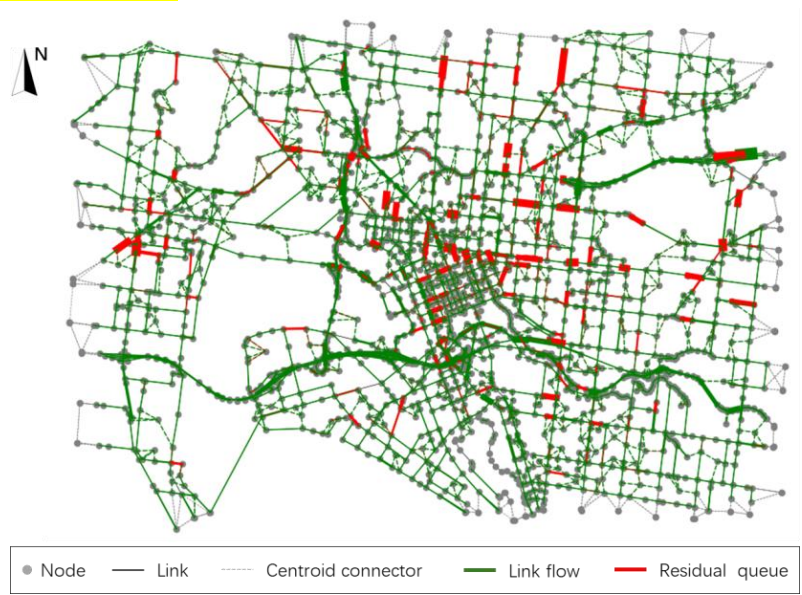

Figure 8. Model results on the real-world road network

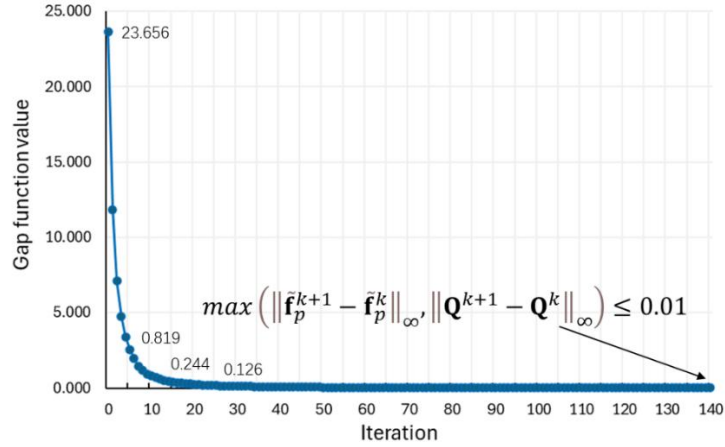

Figure 9. Convergence of the proposed GP-AM algorithm

## 6 Conclusions and further studies

In this paper, a new static traffic assignment model that incorporates the residual queue and queue-dependent link capacity under congested conditions is proposed. To capture the effects of a residual queue on the link capacity, a queue-dependent link capacity is proposed and incorporated into a generalized link cost function. A minimization problem equivalent to the proposed queue-dependent traffic assignment problem is formulated in this paper. Importantly, the uniqueness of the solutions on all links under given traffic conditions (uncongested or congested) is theoretically proved by using a sigmoid function-based reformulation to smooth the objective function.

The numerical results demonstrate that the proposed model can estimate link flows that are smaller than or equal to the link physical capacity under congested conditions. These results are indeed consistent with the observed link flow data from traffic detectors, especially under congested conditions. Importantly, the proposed model is able to produce more reasonable results than traditional models, such that when the traffic demand increases, the equilibrium link flow at the bottleneck first increases and then decreases after reaching the link physical capacity due to the formation of a residual queue. This paper also demonstrates that a conventional static traffic assignment model can be considered a special case of the proposed model with the proposed generalized link cost function. The proposed model yields reasonable estimations of residual queues and queuing delays, which are underestimated by about 30% by the conventional static traffic assignment model with a fixed physical capacity. Furthermore, by varying the parameters in the definition of a residual queue from link flows, the proposed model can account for the effects of different traffic conditions (e.g., a different traffic mix) on the formation of a residual queue.

One of the limitations of the proposed model is that it may produce unrealistic results under severe congestion spillover, where the queue length exceeds the storage capacity of a link. In such cases, the proposed model might underestimate the flow on extremely congested links and overestimate the downstream link flows. Bliemer and Raadsen (2020) and Smith et al. (2019) have demonstrated innovative approaches to incorporate both residual queue and spillback effect in static traffic assignment models, inspiring us to further explore these spatial queue and spillback effects in our proposed model, particularly in larger, more practical networks. It should be noted that the proposed model still belongs to the category of static traffic assignment models without considering time-dependent features. However, it would be valuable to conduct further research on the effects of varying demand over time on network flow and queue dynamics in a dynamic context.

The parameter related to the formation of a residual queue is assumed to be constant in this paper. In practice, this parameter may vary with the link speed so that the residual queue and link flow have various relationships. It merits further studies to calibrate the values of the parameter such as $\gamma$ by empirical validation with observed data. With respect to solution algorithms, it has been noted that the recent development of computational graph may be capable to solve the proposed model more efficiently (Wu et al., 2018; Ma et al., 2020; Kim et al., 2022; Lu et al., 2023; Guarda et al., 2024). It would also be interesting to examine the effects of multiple time periods on the residual queue and traffic assignment models with a queue-dependent capacity (Lawson et al., 1997; Lam et al., 1999; Tampère et al., 2011). Congestion pricing has been considered as one of the most effective methods to alleviate traffic congestion (Yang and Bell, 1997; Meng and Liu, 2012). It merits further studies to investigate how the queuing toll policy affects the proposed generalized link cost function, and the traffic assignment problem.

## 7  Acknowledgments

This research was supported by the National Nature Science Foundation of China (Ref. Nos. 72301023 and 72288101). We would like to thank the editors and anonymous reviewers for their constructive comments to help to improve the paper quality.

## 8  Appendices

**Appendix A**

Table A.1 Summary of different link cost functions in literature

| Category | Literature | Function (remark) |
|---|---|---|
| Flow-based | BPR (1964) | $t = t_f \left[1 + \alpha \left(\frac{v}{C}\right)^\beta\right]$ |
| | Davidson (1966) | $t = t_f \left[1 + j \cdot \frac{v}{C - v}\right]$ |
| | Newell (1982) | $t = t_f \left[1 + \frac{\gamma}{36\mu \cdot t_f} \cdot \left(\frac{D}{\mu}\right)^3\right]$ <br> ($\mu$ is the discharge rate, $\gamma$ is the shape parameter) |
| | Small (1983) | $t = \begin{cases} t_f, \text{if } v \leq C \\ t_f + \frac{1}{2}P \cdot \left(\frac{v}{C} - 1\right), \text{if } v > C \end{cases}$ |
| | Spiess (1990) | $tt = t_f \cdot [2 + \sqrt{\alpha^2(1 - \frac{v}{C})^2 + \beta^2} - \alpha\left(1 - \frac{v}{C}\right) - \beta]$ |
| | Akçelik (1991) | $t = t_f + 0.25T \cdot [\left(\frac{v}{C} - 1\right) + \sqrt{\left(\frac{v}{C} - 1\right)^2 + \frac{8Jv}{TC^2}}]$ |

| Density-based | Kucharski and Drabicki (2017) | $t = t_f \left[1 + \alpha \left(\dfrac{k}{k_c}\right)^\beta\right]$ |
|---|---|---|
| Queue demand-based | Huntsinger and Rouphail (2011) and Moses et al. (2013) | $t = \begin{cases} t_f \left[1 + \alpha \left(\dfrac{v}{C}\right)^\beta\right], & if\ v \le C \\ t_f \left[1 + \alpha \left(\dfrac{D}{C}\right)^\beta\right], & if\ v > C \end{cases}$<br>$D = demand\ at\ capacity + queue$ |
| | Cheng et al. (2022) and Zhou et al. (2022) | $t = t_f \left[1 + \dfrac{\gamma \cdot g(m)}{\mu \cdot t_f} \cdot \left(\dfrac{D}{\mu}\right)^4\right]$<br>(Cubic arrival rates) |
| | Bliemer et al. (2014) and Bliemer and Raadsen (2020) | $t_p = \sum_{a \in A} \dfrac{L_a}{v_f^a} \cdot \delta_{ap} + \tau_p^{queue}$<br>(Separable free-flowing travel time and non-separable path queuing delay) |

## Appendix B

**Transferability of the proposed user equilibrium model to system optimum**

It should be noted that the proposed user equilibrium model with residual queue is able to be directly transferable to a system optimum by considering the total network travel time. The optimization problem of system optimum can be expressed below

$$\min \sum_{a \in \mathbf{A}} (v_a + Q_a) \cdot t_a(v_a, Q_a) \qquad (B.1)$$

subject to

Equations (1), (3), (4), and (6)–(9),

The above optimization problem of system optimum is easily transferable to the user equilibrium-like formulation. By defining $\bar{t}_a(v_a, Q_a) = t_a(v_a, Q_a) + (v_a + Q_a)\dfrac{\partial t_a}{\partial v_a}$ and $\bar{c}_p = \sum_{a \in \boldsymbol{\eta}_p} \delta_{ap} \bar{t}_a \ \forall p \in \mathbf{P}_{rs}, r \in \mathbf{R}, s \in \mathbf{S}$, the KKT conditions can be derived from the Lagrangian function of the system optimum as follows:

$$\hat{f}_p \left(\bar{c}_p + \sum_{a \in \boldsymbol{\eta}_p} \delta_{ap} \mu_a - w_{rs}\right) = 0, \forall p \in \mathbf{P}_{rs}, r \in \mathbf{R}, s \in \mathbf{S} \qquad (B.2a)$$

$$\bar{c}_p + \sum_{a \in \boldsymbol{\eta}_p} \delta_{ap} \mu_a - w_{rs} \ge 0, \forall p \in \mathbf{P}_{rs}, r \in \mathbf{R}, s \in \mathbf{S} \qquad (B.2b)$$

$$D_{rs} = \sum_{p \in \mathbf{P}_{rs}} \hat{f}_p, \forall r \in \mathbf{R}, s \in \mathbf{S} \qquad (B.2c)$$

It can be seen that similar to traditional UE and SO, the only difference between the proposed UE and corresponding SO is the way to calculate link cost and path cost. The link cost in this SO consists of two terms: the former term is the same as the link cost in the proposed UE, while the latter term is the extra cost added to the current flow and residual queue when adding a unit traveling flow.

## Appendix C

**Proof of the equivalency between Equations (10) and (13)**

**Proof:** It can be noticed that the only difference between Equations (10) and (13) is their first terms. Thus, in order to prove the equivalency between Equations (10) and (13), we only need to prove that their first terms are equivalent as follows.

$$\sum_{a \in A}\left[\int_0^{v_a}\left(t_{f,a}+t_{f,a}\beta\left(\frac{x_a}{C_a(Q_a)}\right)^n\right)dx_a\right] = \sum_{a \in A}\left[\int_0^{v_a}\left(t_{f,a}+t_{f,a}\beta\left(\frac{x_a}{C_{a,max}}\right)^{n\varphi^{-Q_a}}\right)dx_a\right]$$

$$\Leftrightarrow \int_0^{v_a}\left(t_{f,a}+t_{f,a}\beta\left(\frac{x_a}{C_a(Q_a)}\right)^n\right)dx_a = \int_0^{v_a}\left(t_{f,a}+t_{f,a}\beta\left(\frac{x_a}{C_{a,max}}\right)^{n\varphi^{-Q_a}}\right)dx_a$$

$$\Leftrightarrow t_{f,a}\beta\left(\frac{x_a}{C_a(Q_a)}\right)^n = t_{f,a}\beta\left(\frac{x_a}{C_{a,max}}\right)^{n\varphi^{-Q_a}} \quad \forall a \in A$$

As $Q_a \geq 0$ according to Equation (2), we can divide the proof into two different cases:

(1) if $Q_a = 0$, for $\forall a \in A$, the left-hand-side term becomes $t_{f,a}\beta\left(\frac{x_a}{C_a(Q_a=0)}\right)^n = t_{f,a}\beta\left(\frac{x_a}{C_{a,max}}\right)^n$, and the right-hand-side term becomes $t_{f,a}\beta\left(\frac{x_a}{C_{a,max}}\right)^{n\varphi^{-Q_a}} = t_{f,a}\beta\left(\frac{x_a}{C_{a,max}}\right)^n$. The equivalency under $Q_a = 0$ is proved.

(2) if $Q_a > 0$, for $\exists a \in A$, then $C_a(Q_a) = v_a$ according to Equation (4). The left-hand-side term becomes $t_{f,a}\beta\left(\frac{x_a}{C_a(Q_a)}\right)^n = t_{f,a}\beta$, and the right-hand-side term becomes $\lim_{\varphi \to +\infty} t_{f,a}\beta\left(\frac{x_a}{C_{a,max}}\right)^{n\varphi^{-Q_a}} = t_{f,a}\beta$. The equivalency under $Q_a > 0$ is proved when the introduced parameter $\varphi$ goes to infinity.

Therefore, the proof of the equivalency between Equations (10) and (13) when the introduced parameter $\varphi$ approaches infinity is completed. It should be noted that the parameter $\varphi$ is normally set to be the Euler's number (i.e., $\varphi = e$), as in the sigmoid function.

## Appendix D

**Sensitivity analysis for parameters $m$ and $n$**

Further sensitivity analysis of parameters $m$ and $n$ has been conducted. Table D.1 demonstrates that as the value of $m$ increases, the link flow on the bottleneck link decreases monotonically, while the residual queue increases. Specifically, parameter $m$ negatively impacts link flow but positively impacts the residual queue on the bottleneck link. However, as shown in Table D.2, parameter $n$ does not have consistent effects on the link flow at the bottleneck, which may be attributed to the occurrence of residual queues.

Table D.1. Sensitivity analysis of the effects of parameter m on results of Link 4 ($\gamma = 0.5, n = 4$)

| $m$ | Link flow (veh/hr) | Residual queue (veh/hr) | Queuing delay (hr) | Traveling time (hr) | Generalized link cost (hr) |
|---|---|---|---|---|---|
| 0.5 | 2395 | 9 | 0.031 | 0.250 | 0.281 |
| 1 | 2350 | 99 | 0.021 | 0.251 | 0.272 |
| 2 | 2276 | 247 | 0.006 | 0.251 | 0.256 |
| 3 | 2253 | 295 | 0.001 | 0.251 | 0.252 |
| 4 | 2248 | 304 | 0.000 | 0.250 | 0.251 |
| 5 | 2247 | 306 | 0.000 | 0.251 | 0.251 |
| 6 | 2247 | 306 | 0.000 | 0.250 | 0.251 |

Table D.2. Sensitivity analysis of the effects of parameter n on results of Link 4 ($\gamma = 0.5, m = 1$)

| $n$ | Link flow (veh/hr) | Residual queue (veh/hr) | Queuing delay (hr) | Traveling time (hr) | Generalized link cost (hr) |
|---|---|---|---|---|---|
| 0.5 | 2244 | 0 | 0.000 | 0.248 | 0.248 |
| 1 | 2309 | 0 | 0.000 | 0.247 | 0.247 |
| 2 | **2390** | 0 | 0.000 | 0.250 | 0.250 |
| 3 | 2372 | 57 | 0.012 | 0.251 | 0.262 |
| 4 | 2350 | 99 | 0.021 | 0.251 | 0.272 |
| 5 | 2338 | 124 | 0.026 | 0.251 | 0.277 |
| 6 | 2332 | 137 | 0.029 | 0.251 | 0.280 |
| 7 | 2328 | 143 | 0.031 | 0.251 | 0.281 |
| 8 | 2327 | 145 | 0.031 | 0.251 | 0.282 |

# 9   References

Akçelik, R., 1991. Travel time functions for transport planning purposes: Davidson's function, its time dependent form and alternative travel time function. *Australian Road Research* 21.

Akçelik, R., Rouphail, N.M., 1993. Estimation of delays at traffic signals for variable demand conditions. *Transportation Research Part B: Methodological* 27, 109-131.

Beck, A., 2015. On the convergence of alternating minimization for convex programming with applications to iteratively reweighted least squares and decomposition schemes. *SIAM Journal on Optimization* 25, 185-209.

Beckmann, M., McGuire, C.B., Winsten, C.B., 1956. Studies in the Economics of Transportation. Yale University Press, New Haven.

Belezamo, B., 2020. Data-driven Methods for Characterizing Transportation System Performances Under Congested Conditions: A Phoenix Study. Arizona State University.

Bell, M.G., Iida, Y., 1997. Transportation network analysis. John Wiley & Sons, Chichester, U.K.

Bliemer, M.C.J., Raadsen, M.P.H., Smits, E.S., Zhou, B., Bell, M.G.H., 2014. Quasi-dynamic traffic assignment with residual point queues incorporating a first order node model. *Transportation*

*Research Part B: Methodological* 68, 363-384.

Bliemer, M.C.J., Raadsen, M.P.H., Brederode, L.J.N., Bell, M.G.H., Wismans, L.J.J., Smith, M.J., 2017. Genetics of traffic assignment models for strategic transport planning. *Transport Reviews* 37, 56-78

Bliemer, M.C.J., Raadsen, M.P.H., 2020. Static traffic assignment with residual queues and spillback. *Transportation Research Part B: Methodological* 132, 303-319.

BPR, 1964. Traffic assignment manual. *Presented at the Planning Division, US Department of Commerce, Washington DC.*

Cheng, Q., Liu, Z., Lin, Y., Zhou, X.S., 2021. An s-shaped three-parameter (S3) traffic stream model with consistent car following relationship. *Transportation Research Part B: Methodological* 153, 246-271.

Cheng, Q., Liu, Z., Guo, J., Wu, X., Pendyala, R., Belezamo, B., Zhou, X.S., 2022. Estimating key traffic state parameters through parsimonious spatial queue models. *Transportation Research Part C: Emerging Technologies* 137, 103596.

Cheng, Q., Lin, Y., Zhou, X.S., Liu, Z., 2023. Analytical formulation for explaining the variations in traffic states: A fundamental diagram modeling perspective with stochastic parameters. *European Journal of Operational Research*.

Davidson, K., 1966. A flow travel time relationship for use in transportation planning, *Australian Road Research Board (ARRB) Conference, 3rd, 1966, Sydney*.

De Palma, A., Lindsey, R., 2002. Comparison of morning and evening commutes in the Vickrey bottleneck model. *Transportation Research Record* 1807, 26-33.

Du, L., Gong, S., 2016. Stochastic Poisson game for an online decentralized and coordinated parking mechanism. *Transportation Research Part B: Methodological* 87, 44-63

Fu, H., Lam, W.H.K., Shao, H., Ma, W., Chen, B.Y., Ho, H.W., 2022. Optimization of multi-type sensor locations for simultaneous estimation of origin-destination demands and link travel times with covariance effects. *Transportation Research Part B: Methodological* 166, 19-47.

Guarda, P., Battifarano, M., Qian, S., 2024. Estimating network flow and travel behavior using day-to-day system-level data: A computational graph approach. *Transportation Research Part C: Emerging Technologies* 158, 104409.

Han, K., Friesz, T.L., Yao, T., 2013a. A partial differential equation formulation of Vickrey's bottleneck model, part I: Methodology and theoretical analysis. *Transportation Research Part B: Methodological* 49, 55-74.

Han, K., Friesz, T.L., Yao, T., 2013b. A partial differential equation formulation of Vickrey's bottleneck model, part II: Numerical analysis and computation. *Transportation Research Part B: Methodological* 49, 75-93.

HCM, 2010. Highway capacity manual. *Washington, DC* 2, 1.

Huntsinger, L.F., Rouphail, N.M., 2011. Bottleneck and queuing analysis: Calibrating volume–delay functions of travel demand models. *Transportation research record* 2255, 117-124.

Kim, T., Zhou, X., Pendyala, R.M., 2022. Computational graph-based framework for integrating econometric models and machine learning algorithms in emerging data-driven analytical environments. *Transportmetrica A: Transport Science* 18, 1346-1375.

Kucharski, R., Drabicki, A., 2017. Estimating macroscopic volume delay functions with the traffic density derived from measured speeds and flows. *Journal of Advanced Transportation* 2017.

Lam, W.H.K., Zhang, Y., Yin, Y., 1999. Time-dependent model for departure time and route choices in networks with queues. *transportation research record* 1667, 33-41.

Lam, W.H.K., Zhang, Y., 2000. Capacity-constrained traffic assignment in networks with residual queues. *Journal of Transportation Engineering* 126, 121–128.

Lawson, T.W., Lovell, D.J., Daganzo, C.F., 1997. Using input-output diagram to determine spatial and temporal extents of a queue upstream of a bottleneck. *Transportation Research Record* 1572, 140-147.

LeBlanc, L.J., Morlok, E.K., Pierskalla, W.P., 1975. An efficient approach to solving the road network equilibrium traffic assignment problem. *Transportation Research* 9, 309-318

Li, G., Chen, A., 2023. Strategy-based transit stochastic user equilibrium model with capacity and number-of-transfers constraints. *European Journal of Operational Research* 305, 164-183.

Li, J., Zhang, H.M., 2015. A generalized queuing model and its solution properties. *Transportation Research Part B: Methodological* 79, 78-92.

Li, Z.-C., Huang, H.-J., Yang, H., 2020. Fifty years of the bottleneck model: A bibliometric review and future research directions. *Transportation research part B: methodological* 139, 311-342.

Lo, H.K., Szeto, W.Y., 2005. Road pricing modeling for hyper-congestion. *Transportation Research Part A: Policy and Practice* 39, 705-722.

Lu, J., Li, C., Wu, X.B., Zhou, X.S., 2023. Physics-informed neural networks for integrated traffic state and queue profile estimation: A differentiable programming approach on layered computational graphs. *Transportation Research Part C: Emerging Technologies* 153, 104224.

Lu, J., Zhou, X.S., 2023. Virtual track networks: A hierarchical modeling framework and open-source tools for simplified and efficient connected and automated mobility (CAM) system design based on general modeling network specification (GMNS). *Transportation Research Part C: Emerging Technologies* 153, 104223.

Ma, W., Qian, Z., 2017. On the variance of recurrent traffic flow for statistical traffic assignment. *Transportation Research Part C: Emerging Technologies* 81, 57-82.

Ma, W., Qian, Z., 2018. Statistical inference of probabilistic origin-destination demand using day-to-day traffic data. *Transportation Research Part C: Emerging Technologies* 88, 227-256.

Ma, W., Pi, X., Qian, S., 2020. Estimating multi-class dynamic origin-destination demand through a forward-backward algorithm on computational graphs. *Transportation Research Part C: Emerging Technologies* 119, 102747.

Meng, Q., Liu, Z., 2012. Impact analysis of cordon-based congestion pricing on mode-split for a bimodal transportation network. *Transportation Research Part C: Emerging Technologies* 21, 134-147.

Moses, R., Mtoi, E., Ruegg, S., McBean, H., Brinckerhoff, P., 2013. Development of speed models for improving travel forecasting and highway performance evaluation. Florida. Dept. of Transportation.

Nesterov, Y., 2018. *Lectures on convex optimization*. Springer.

Newell, C., 1982. *Applications of queueing theory*. Springer Netherlands, Dordrecht. .

Newell, G.F., 1968a. Queues with time-dependent arrival rates I—the transition through saturation. *Journal of Applied Probability* 5, 436-451.

Newell, G.F., 1968b. Queues with time-dependent arrival rates. II—The maximum queue and the return to equilibrium. *Journal of Applied Probability* 5, 579-590.

Newell, G.F., 1968c. Queues with time-dependent arrival rates. III—A mild rush hour. *Journal of Applied Probability* 5, 591-606.

Nie, Y.M., Zhang, H.M., Lee, D.H., 2004. Models and algorithms for the traffic assignment problem with link capacity constraints. *Transportation Research Part B: Methodological* 38, 285-312.

Nourmohammadi, F., Mansourianfar, M., Shafiei, S., Gu, Z., Saberi, M., 2021. An Open GMNS Dataset of a Dynamic Multi-Modal Transportation Network Model of Melbourne, Australia. *Data* 6, 21.

Ordóñez, F., Stier-Moses, N.E., 2010. Wardrop equilibria with risk-averse users. *Transportation Science* 44, 63-86.

Pan, Y., Guo, J., Chen, Y., 2022. Calibration of dynamic volume-delay functions: A rolling horizon-based parsimonious modeling perspective. *Transportation research record* 2676, 606-620.

Pan, Y.A., Zheng, H., Guo, J., Chen, Y., 2023. Modified volume-delay function based on traffic fundamental diagram: A practical calibration framework for estimating congested and uncongested conditions. *Journal of transportation engineering, Part A: Systems* 149, 04023112.

Pan, Y.A., Guo, J., Chen, Y., Abbasi, M., List, G., Zhou, X.S., 2024. A Review of Volume-Delay Functions: Connecting Theoretical Fundamental, Practical Deployment and Emerging Applications. *Available at SSRN: https://ssrn.com/abstract=4143355*.

Raadsen, M.P., Bliemer, M.C., 2019a. Steady-state link travel time methods: Formulation, derivation, classification, and unification. *Transportation Research Part B: Methodological* 122, 167-191.

Raadsen, M.P.H., Bliemer, M.C.J., 2019b. Steady-state link travel time methods: Formulation, derivation, classification, and unification. *Transportation Research Part B: Methodological* 122, 167-191

Ringhand, M., Siebke, C., Bäumler, M., Petzoldt, T., 2022. Approaching intersections: Gaze behavior of drivers depending on traffic, intersection type, driving maneuver, and secondary task involvement. *Transportation research part F: traffic psychology and behaviour* 91, 116-135.

Rusnock, P., Kerr-Lawson, A., 2005. Bolzano and uniform continuity. *Historia Mathematica* 32, 303-311.

Shao, H., Lam, W.H.K., Tam, M.L., 2006. A reliability-based stochastic traffic assignment model for network with multiple user classes under uncertainty in demand. *Networks and Spatial Economics* 6, 173-204.

Shao, H., Lam, W.H.K., Sumalee, A., Chen, A., 2013. Journey time estimator for assessment of road network performance under demand uncertainty. *Transportation Research Part C: Emerging Technologies* 35, 244-262.

Shao, H., Lam, W.H.K., Sumalee, A., Chen, A., Hazelton, M.L., 2014. Estimation of mean and covariance of peak hour origin–destination demands from day-to-day traffic counts. *Transportation Research Part B: Methodological* 68, 52-75

Shao, H., Lam, W.H.K., Sumalee, A., Hazelton, M.L., 2015. Estimation of mean and covariance of stochastic multi-class OD demands from classified traffic counts. *Transportation Research Part C: Emerging Technologies* 7, 192-211.

Sheffi, Y., 1985. Urban transportation networks : Equilibrium analysis with mathematical programming methods. Prentice-Hall, Englewood Cliffs, NJ, USA.

Sheu, J.-B., Pan, C., 2014. A method for designing centralized emergency supply network to respond to large-scale natural disasters. *Transportation Research Part B: Methodological* 67, 284-305

Small, K., Verhoef, E.T., 2007. The economics of urban transportation. Routledge.

Small, K.A., 1983. The incidence of congestion tolls on urban highways. *Journal of urban economics* 13, 90-111.

Smith, M., Huang, W., Viti, F., Tampère, C.M.J., Lo, H.K., 2019. Quasi-dynamic traffic assignment with spatial queueing, control and blocking back. *Transportation Research Part B: Methodological* 122, 140-166

Spiess, H., 1990. Conical volume-delay functions. *Transportation Science* 24, 153-158.

Tampère, C.M., Corthout, R., Cattrysse, D., Immers, L.H., 2011. A generic class of first order node models for dynamic macroscopic simulation of traffic flows. *Transportation Research Part B: Methodological* 45, 289-309.

Transport Department, 2023. The Third Comprehensive Transport Study : Final Report https://www.td.gov.hk/en/publications_and_press_releases/publications/free_publications/the_third_comprehensive_transport_study/index.html.

Tseng, P., 1991. Applications of a splitting algorithm to decomposition in convex programming and variational inequalities. *SIAM Journal on Control and Optimization* 29, 119-138

Vickrey, W.S., 1969. Congestion theory and transport investment. *The American economic review* 59, 251-260.

Wardrop, J.G., 1952. Road paper. some theoretical aspects of road traffic research. *Proceedings of the institution of civil engineers* 1, 325-362.

Wong, C., Wong, S., Lo, H.K., 2007. Reserve capacity of a signal-controlled network considering the effect of physical queuing. *Transportation and Traffic Theory*, 533-553.

Wu, X., Guo, J., Xian, K., Zhou, X., 2018. Hierarchical travel demand estimation using multiple data sources: A forward and backward propagation algorithmic framework on a layered computational graph. *Transportation Research Part C: Emerging Technologies* 96, 321-346.

Wu, X.B., Dutta, A., Zhang, W., Zhu, H., Livshits, V., Zhou, X.S., 2021. Characterization and calibration of volume-to-capacity ratio in volume-delay functions on freeways based on a queue analysis approach. *Transportation Research Record: Journal of the Transportation Research Board*.

Xie, J., Nie, Y., 2019. A new algorithm for achieving proportionality in user equilibrium traffic assignment. *Transportation science* 53, 566-584.

Xu, Z., Chen, Z., Yin, Y., Ye, J., 2021. Equilibrium analysis of urban traffic networks with ride-sourcing services. *Transportation science* 55, 1260-1279.

Xu, Z., Xie, J., Liu, X., Nie, Y., 2022. Hyperbush algorithm for strategy-based equilibrium traffic assignment problems. *Transportation science* 56, 877-903.

Xu, Z., Chen, A., Liu, X., 2023. Time and toll trade-off with heterogeneous users: A continuous time surplus maximization bi-objective user equilibrium model. *Transportation Research Part B: Methodological* 173, 31-58.

Xu, Z., Peng, Y., Li, G., Chen, A., Liu, X., 2024. Range-constrained traffic assignment for electric vehicles under heterogeneous range anxiety. *Transportation Research Part C: Emerging Technologies* 158, 104419.

Yang, H., Yagar, S., 1995. Traffic assignment and signal control in saturated road networks. *Transportation Research Part A: Policy and Practice* 29, 125-139.

Yang, H., Bell, M.G., 1997. Traffic restraint, road pricing and network equilibrium. *Transportation Research Part B: Methodological* 31, 303-314.

Zhang, H., Nie, Y., Qian, Z., 2013. Modelling network flow with and without link interactions: the cases of point queue, spatial queue and cell transmission model. *Transportmetrica B: Transport Dynamics* 1, 33-51.

Zhou, X.S., Cheng, Q., Wu, X., Li, P., Belezamo, B., Lu, J., Abbasi, M., 2022. A meso-to-macro cross-resolution performance approach for connecting polynomial arrival queue model to volume-delay function with inflow demand-to-capacity ratio. *Multimodal Transportation* 1, 100017.